\documentclass[journal]{IEEEtran}

\usepackage[numbers,sort&compress,square]{natbib}
\usepackage{comment}
\usepackage{hyperref}
\usepackage{amsmath,amssymb,amsfonts}
\usepackage{upgreek}
\usepackage{graphicx}
\usepackage{textcomp}
\usepackage{xcolor}
\usepackage{tabu,multirow}
\usepackage{mathtools}
\usepackage{tabularx}
\usepackage{ragged2e} 
\usepackage{stfloats}

\def\BibTeX{{\rm B\kern-.05em{\sc i\kern-.025em b}\kern-.08em
    T\kern-.1667em\lower.7ex\hbox{E}\kern-.125emX}}

\newcommand{\quotes}[1]{``#1''}

\newcolumntype{Y}{>{\centering\arraybackslash}X}

\begin{document}

\title{Vehicular Visible Light Positioning for \\Collision Avoidance and Platooning: A Survey\\}

\author{Burak~Soner,~\IEEEmembership{Member,~IEEE,}
	Merve~Karakaş,~\IEEEmembership{Student Member,~IEEE,}
	Utku~Noyan,~\IEEEmembership{Student Member,~IEEE,}
	Furkan~Şahbaz,~\IEEEmembership{Student Member,~IEEE,}	and~Sinem~Coleri,~\IEEEmembership{Senior Member,~IEEE,}
\thanks{Burak Soner, Merve Karakaş, Utku Noyan, Furkan Şahbaz and Sinem Coleri are with the Department of Electrical and Electronics Engineering, Koc University, 34450 Istanbul, Turkey (e-mail: \{bsoner16, mkarakas16, unoyan16, fsahbaz16, scoleri\}@ku.edu.tr). The authors acknowledge the support of CHIST-ERA grant CHISTERA-18-SDCDN-001, the Scientific and Technological Council of Turkey 119E350 and Ford Otosan.}}

\maketitle

\begin{abstract}

Relative vehicle positioning methods can contribute to safer and more efficient autonomous driving by enabling collision avoidance and platooning applications. For full automation, these applications require cm-level positioning accuracy and greater than 50 Hz update rate. Since sensor-based methods (e.g., LIDAR, cameras) have not been able to reliably satisfy these requirements under all conditions so far, complementary methods are sought. Recently, {\color{black}positioning based on visible light communication signals from vehicle head/tail LED lights (VLP) has shown significant promise as a complementary method} attaining cm-level accuracy and near-kHz rate in realistic driving scenarios. {\color{black}Vehicular VLP methods measure relative bearing (angle) or range (distance) of transmitters (i.e., head/tail lights) based on received signals from on-board photodiodes and estimate transmitter relative positions based on those measurements.} In this survey, {\color{black}we first review existing vehicular VLP methods and propose a new method that advances the state-of-the-art in positioning performance. Next, we analyze the theoretical and simulated performance of all methods in realistic driving scenarios under challenging noise and weather conditions, real asymmetric light beam patterns and different vehicle dimensions and light placements. Our simulation results show that the newly proposed VLP method is the overall best performer, and can indeed satisfy the accuracy and rate requirements for localization in collision avoidance and platooning applications within practical constraints.} Finally, we discuss remaining open challenges that are faced for the deployment of VLP solutions in the automotive sector and further research questions.

\end{abstract}


\section{Introduction}

Collision avoidance and platooning are essential applications for safe and efficient autonomous driving \cite{ford_trust}. For full autonomy, these applications require vehicle-to-vehicle relative localization with cm-level accuracy and greater than 50 Hz rate \cite{caveney, shladover}. Conventional relative vehicle localization methods using repurposed sensor-based object trackers (e.g., LIDAR, RADAR, cameras) and \quotes{co-operative} schemes based on sharing GPS information {\color{black}have shown promise on this front in recent years, especially with the advent of deep learning based methods. However, these methods fail to meet all stringent requirements simultaneously due to sensor rate and accuracy limitations or high computational complexity, necessitating complementary solutions. Moreover, the cost associated with sensor-based systems are getting increasingly prohibitive since they require powerful accelerators such as GPUs to process high-dimensional data (e.g., billions of pixels for cameras) \cite{depontemuller, guney_cvav, lidar_ref}. On the other hand,} communication-based methods, which use {\color{black}only 1D received signals over time for localizing antennas on vehicles, fundamentally promise the cm-level accuracy at much greater than 50 Hz rate on low cost processors thanks to their low computational complexity and robust target detection procedures \cite{vvlc_survey_memedi, cellularDsrcInterwork, coop_localization, balico_survey, soner_tvt}. For this reason, communication-based methods can be a suitable complementary solution.}

Communication-based relative vehicle localization methods have primarily used radio frequency (RF) and visible light communication (VLC) signals for positioning. However, previous studies on localization with RF communications like the IEEE 802.11p based ITS G5 and IEEE 1609 WAVE \cite{etsi_wave}, and the cellular-based LTE and C-V2X \cite{etsi_cv2x} show that their accuracy is limited to 1-10 m in realistic use cases. This is primarily due to severe congestion, multi-path interference (fueled by non-directional antenna patterns) and tight synchronization requirements in RF \cite{liu_wiComPosSurvey, vehaoa, alam_53, NAlam_cooperative, hybrid_globecom}. On the other hand, vehicular visible light positioning (VLP), which uses received directional VLC signals from modulated vehicle head/tail LED lights for positioning, fundamentally promises the required cm-level accuracy owing to the dominant line-of-sight (LoS) propagation characteristics of the channel at distances relevant to collision avoidance and platooning \cite{roadmap_vlp, inoutdoor_survey_zhuang, vlc_channel_lee, vlc_channel_wnl, vlc_channel_wnl_ml}. {\color{black}Therefore, vehicular VLP is closer to realizing the promises of communication-based positioning compared to RF.}

{\color{black}Fundamental technologies in the field of vehicular VLP were first conceived for indoor VLC and VLP, which consider localization of RX mobile devices under TX LED ceiling lights \cite{mfkeskin_proceeding, inoutdoor_survey_zhuang, luo2017indoor, indoor_survey_hopdo, mfkeskin_ranging}. However, vehicular VLC/VLP emerged as a separate field due to three major differences between the indoor and vehicular domains \cite{vvlc_survey_memedi, caileanSurvey_vlcAutoChlgs}:}

\begin{enumerate}
	\vspace{1mm}
	\item \textit{{\color{black}Much fewer TX units}}: LED head/tail lights {\color{black}and co-located photodiodes are the natural choice for vehicular VLC TX/RX units. Therefore, vehicular use cases consider only two head or two (max. three \cite{turan_3tx}) tail lights as TX/RX units. Indoor use cases require many more TXs (ceiling lights) for high accuracy \cite{teamyuksel_leastsqrs, adoa_3d_vlp, jung_tdoa, nadeem_tdoa, yang_aoa_rss, tdclittle_hybrid, mazuelas_softlocalization}.}
	\vspace{2mm}
	\item \textit{Very high mobility}: Vehicular use cases have very high TX-RX relative mobility resulting in frequent fluctuations in {\color{black}SNR.} On the other hand, indoor use cases {\color{black}typically rely on very low RX mobility and static parallel TXs (ceiling lights) for high accuracy (e.g., long-interval averaging at the cost of lower rate) \cite{mfkeskin_proceeding, heidi_crb_aoa, heidi_3d_aoa}.}
	\vspace{2mm}
	\item \textit{LoS-dominant channel}: The vehicular VLC channel is LoS-dominant \cite{vlc_channel_wnl, vlc_channel_wnl_ml, vlc_survey} but indoor channels typically contain significant non-LoS components {\color{black}due to random reflections from many surfaces. For this reason, while vehicular VLP achieves higher accuracy with geometric algorithms that exploit the strong correlation between TX-RX bearing/range and the RX signal characteristics, indoor methods achieve higher accuracy from statistical inference methods like direct estimation \cite{heidi_singleStep, mfkeskin_bounds} which estimate position directly from RX signals without explicitly measuring bearing/range.}
	\vspace{1mm}
\end{enumerate}

\begin{table*}[b]
	\renewcommand{\arraystretch}{1.5}
	\caption{Existing VLP Methods, Indoor vs. Vehicular}
	\label{existing_vlp}
	\centering
	\begin{tabu}{c | p{0.12\linewidth} | p{0.12\linewidth} | p{0.5\linewidth}}
		reference & indoor/vehicular & estimation method & useful for collision avoidance and platooning? \\
		\tabucline[1.2pt]{1-4}
		\cite{heidi_singleStep} & \multirow{6}{\linewidth}{indoor} & direct & \multirow{4}{\linewidth}{$\times$, not applicable, requires more TXs than available on the target vehicle and perfect alignment between RX and TXs. \cite{mfkeskin_proceeding} summarizes all methods in this category. } \\
		\cline{1-1} \cline{3-3}
		\cite{heidi_crb_aoa, teamyuksel_leastsqrs, adoa_3d_vlp} & & bearing-based & \\
		\cline{1-1} \cline{3-3}
		\cite{jung_tdoa, nadeem_tdoa}& & range-based &  \\
		\cline{1-1} \cline{3-3}
		\cite{yang_aoa_rss, tdclittle_hybrid}& & hybrid & \\
		\cline{1-1} \cline{3-4}
		\cite{heidi_3d_aoa}& & bearing-based & \multirow{2}{\linewidth}{$\times$, not applicable, algorithms diverge in the high-mobility case without good initialization and/or training data } \\
		\cline{1-1} \cline{3-3}
		\cite{mazuelas_softlocalization}& & range-based & \\
		\hline
		\hline
		\cite{mahmoud_cokyeni_v2i_ann} & \multirow{6}{\linewidth}{vehicular} & direct & \multirow{3}{\linewidth}{$\times$, vehicular but not applicable since fixed co-planar TX arrays are assumed (traffic/road lights); these methods are useful for vehicle-to-infrastructure applications} \\
		\cline{1-1} \cline{3-3}
		\cite{shieh_v2i_pyramid, yamazato2014_camAoA} & & bearing-based & \\
		\cline{1-1} \cline{3-3}
		\cite{bai_tdoa, he_cokyeni_camera_v2i} & & range-based & \\
		\cline{1-1} \cline{3-4}
		\cite{ir_bizimgeo, ir_bizimgeo2, shieh_v2v} & & bearing-based & \checkmark, but low-precision positioning due to the use of tilted (pyramid) RX units. \\
		\cline{1-1} \cline{3-4}
		\cite{yamazato2014_hispeed} & & bearing-based & \checkmark, but low VLC rate (500 bps/link \cite{caileanSurvey_vlcAutoChlgs, ziehn2020imaging}) and costly (high-FPS camera).\\
		\cline{1-1} \cline{3-4}
		\cite{xu_matchedfilt_lm} & & range-based & \checkmark, but needs $\gg$2 RXs for high accuracy. \\
		\hline
		\hline
		{\color{black}\cite{soner_pimrc, soner_tvt, techreport_vlp_sonercoleri}} & \multirow{2}{\linewidth}{{\color{black}vehicular}} & {\color{black}bearing-based} & \multirow{2}{\linewidth}{\checkmark \checkmark, {\bf {\color{black}state-of-the-art methods for vehicle collision avoidance and platooning, analyzed in this paper}}}\\
		\cline{1-1} \cline{3-3}
		{\color{black}\cite{roberts_pdoa, becha_positioning, techreport_vlp_sonercoleri}} & & {\color{black}range-based} & \\
		\hline
	\end{tabu}
	\vspace{1mm}
\end{table*}

{\color{black}We categorize existing major works in vehicular VLP as well as indoor methods in Table \ref{existing_vlp} and discuss how the current state-of-the-art in vehicular VLP is distinguished from others for historical reasons. Vehicle-to-infrastructure VLP methods that consider road/traffic lights as TX units are proposed as an adaptation of indoor VLP methods \cite{bai_tdoa, ann_vlp_roadside, mahmoud_cokyeni_v2i_ann, shieh_v2i_pyramid, yamazato2014_camAoA, he_cokyeni_camera_v2i}. Although these methods enable general traffic information and road safety applications in urban settings, they} are not suitable {\color{black}for collision avoidance and platooning applications} due to low availability of road lighting. {\color{black}Vehicle-to-vehicle VLP using head/taillights is considered as a solution, and suitable RX architectures are proposed accordingly.} Tilted (pyramidal) photodiodes based \cite{ir_bizimgeo, ir_bizimgeo2, shieh_v2v} and camera-based methods \cite{yamazato2014_hispeed} are proposed, but these are costly, they provide low precision position estimation and prohibitively low VLC rate. {\color{black}The state-of-the-art in vehicular VLP rather has lower cost units ($<$20 USD \cite{soner_tvt, soner2020low}) capable of high precision measurement and do not negatively affect VLC rate.}

\vspace{2mm}

Current state-of-the-art vehicular VLP methods consider {\color{black}either direct range measurements \cite{becha_positioning} or direct bearing measurements \cite{soner_tvt} from two receivers, differential versions of those \cite{roberts_pdoa, techreport_vlp_sonercoleri} or simply two time-consecutive measurements from a single receiver to obtain a running fix \cite{soner_pimrc, techreport_vlp_sonercoleri}. These existing simple vehicular VLP methods all utilize two-step estimation with only two TXs and two RXs at most on the vehicles}. First, received noisy VLC signals are processed for measuring physical system parameters, i.e., relative bearing (angle-of-arrival, AoA) or range (distance) between a TX and one or more receivers (RX), with a certain level of accuracy. Next, these individual parameter measurements are combined to estimate the position of the VLC TX relative to the VLC RXs. {\color{black}There are two main sources of positioning error for these methods: 1) Low signal-to-noise ratio (SNR) on the received VLC signal, and 2) high vehicle mobility. Low SNR causes statistical error and occurs either due to high geometric signal attenuation (e.g., long distance), or due to high noise (e.g., high shot noise due to sunlight). High vehicle mobility causes a deterministic error (i.e., bias) and} occurs due to the finite rate and latency of the position estimations, i.e., estimations cannot \quotes{catch up} with the actual position when the relative {\color{black}vehicle speeds (between TX-RX) are high. This fundamental error affects all finite-rate localization methods \cite{trj_pred, prob_trj_pred} and decreases with higher positioning rate (greater than 50 Hz rate is required for typical driving scenarios \cite{soner_tvt})}. For this reason, the primary design goal of vehicular VLP {\color{black}is providing cm-level accuracy at greater than 50 Hz positioning rate despite low received SNR and high mobility.}

{\color{black}The existing two-step vehicular VLP methods have already shown promising performance, but there is still room for improvement. Specifically, while range-based positioning methods excel in longitudinal positioning, bearing-based ones perform lateral positioning with higher accuracy, and their combination has not yet been explored \cite{techreport_vlp_sonercoleri}. Furthermore, comparative performance benchmarks for vehicular VLP methods under both realistic and challenging road and weather conditions as well as practical constraints such as different vehicle dimensions and realistic automotive light beam patterns, are not available in the literature.}

\vspace{2mm}

In this paper, we survey the state-of-the-art in vehicular VLP, {\color{black}propose a novel vehicular VLP method that advances the state-of-the-art}, analyze the theoretical and simulated performances of all relevant methods for collision avoidance and platooning applications, and discuss the remaining open questions and challenges for widespread deployment of vehicular VLP. Our main contributions are summarized as follows:

\vspace{1mm}
\begin{itemize}
	\item We review feasible bearing/range measurement techniques and the existing algorithms that use them for position estimation in vehicular VLP for the first time in the literature.
	\vspace{2mm}
	\item {\color{black}We propose a new method that advances the state-of-the-art in vehicular VLP by combining bearing and range measurements. The method utilizes the fact that higher accuracy is obtained from bearing-based methods for the lateral axis and from range-based methods for the longitudinal axis, to provide optimal performance.}
	\vspace{2mm} 
	\item We benchmark the state-of-the-art in vehicular VLP performance by simulating all methods under the same realistic driving scenarios, for the first time in the literature. {\color{black}While earlier studies have analyzed individual vehicular VLP methods \cite{soner_tvt, becha_positioning, smartAutoLighting}, we improve upon these by providing extended fair comparison results under recorded collision course trajectories, high mobility scenarios, measured real asymmetric light patterns and different vehicle dimensions. Moreover, we evaluate the associated Cramer-Rao Lower Bounds (CRLBs) for each algorithm (derived in \cite{techreport_vlp_sonercoleri}) under realistic driving scenarios to bridge the gap between theoretical and simulated performance.} We provide the complete source code in GitHub \footnote{ \href{https://github.com/sonebu/vehicular-vlp-simulations}{https://github.com/sonebu/vehicular-vlp-simulations}} to motivate further benchmarks on the topic.
	\vspace{2mm}
	\item We discuss the open research questions and challenges faced for the widespread deployment of vehicular VLP in the automotive sector and elaborate on potential solutions in detail, for the first time in the literature.
\end{itemize}
\vspace{1mm}

\begin{figure}[t]
	\centering
	\includegraphics[width=0.50\textwidth]{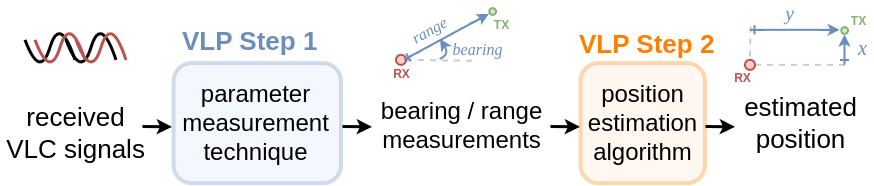}
	\vspace{-1mm}
	\caption{Anatomy of two-step vehicular VLP methods using bearing / range measurements for position estimation.}
	\vspace{-1mm}
	\label{anatomy}
\end{figure}

The rest of the paper is organized as follows. The vehicular VLP system model and the mathematical definition of the received VLC signals are described in Section II. Bearing/range measurement techniques and existing VLP algorithms using those techniques are reviewed, {\color{black}and the newly proposed hybrid positioning algorithm is presented in Section III}. The simulations under realistic collision avoidance and platooning scenarios are presented in Section IV. Remaining open questions and challenges are discussed in Section V. Our conclusions are presented in Section VI.

\section{System Description}

\begin{figure}[t]
	\centering
	\includegraphics[width=0.48\textwidth]{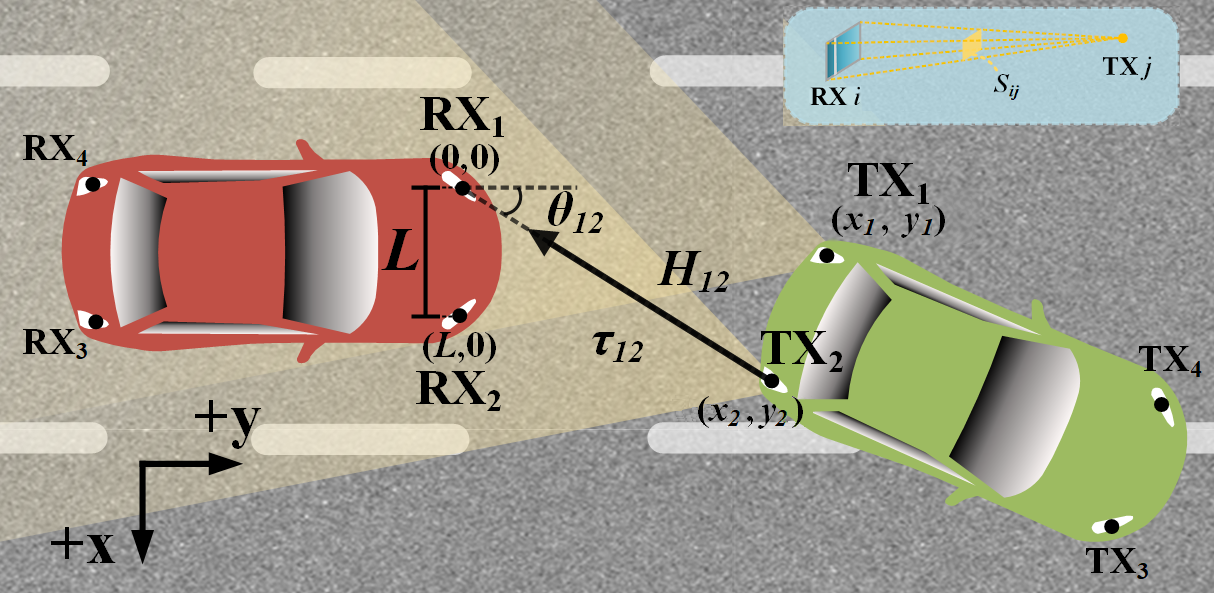}
	\vspace{2mm}
	\caption{System model: The ego vehicle (red) estimates the 2D relative positions of two TXs on the target vehicle (green), i.e., $(x_{1}, y_{1})$ and $(x_{2}, y_{2})$, for vehicle localization. Channel characteristics between TX 2 and RX 1 are depicted as an example: $H_{12}$ is the channel gain, $\tau_{12}$ is the propagation delay and $\theta_{12}$ is the AoA from TX 2 to RX 1. $S_{ij}$ is the angle subtended by RX $i$ with respect to point source TX $j$, and $L$ is the RX separation.}
	\vspace{-2mm}
	\label{sysmdl}
\end{figure}

This section describes the vehicular VLP system model assumptions, presents the problem definition for relative vehicle localization using parameter measurements (i.e., two-step estimation as shown in Fig. \ref{anatomy}), and provides the mathematical model of the received VLC signals that are used for those measurements. The system model is depicted in Fig. \ref{sysmdl}.


\subsection{System Model Assumptions}
\vspace{1mm}
The model is based on the following assumptions (\textit{A\#}):
\begin{itemize}
	\vspace{3mm}
	\item \textit{A1:} Vehicles cruise on piecewise-flat roads, i.e., their pitch angles with the horizon is the same. This assumption, which reduces the 3D positioning problem to 2D (i.e., the road plane), is reasonable for collision avoidance and platooning scenarios where vehicles are within \linebreak1-20 m of each other driving at $\geq$30 km/h \cite{flatroad_ref1,flatroad_ref3}.	
	\vspace{2mm}
	\item \textit{A2:} Vehicles contain VLC units with LED TXs \footnote{TX design {\color{black}(both power as well as the 3D asymmetrical beam pattern \cite{vvlc_survey_memedi})} is subject to strict traffic and eye safety regulations \cite{txreg_iec62471, txreg_nhtsa, txreg_unece_r112, twoheadlight_nhtsa}.} and photodiode-based RXs \footnote{RX design is not constrained: Various front-end designs that explore different optical and electrical trade-offs exist \cite{ada_receivers, vvlc_survey_memedi, goncalves_survey, soner_tvt, kahn1998imaging, rx_bluefilter}.} on their head and tail lights (i.e., 4 VLC units in total), sustaining reliable LoS communication. {\color{black}A minimal setup (2 RXs on each face) is assumed to assess worst case performance.}
	\vspace{2mm}
	\item \textit{A3}: VLC TX units are assumed to be point sources \footnote{{\color{black}Note that this only implies that they are not extended sources. No assumption is made about the shape of the radiation pattern.}}. This assumption holds for the intended applications since the photometric distance for vehicle LED head/tail lights (a maximum of approximately 50 cm \cite{iala_farfield, autoOptikDist_farfield, 5to15_farfield}), is smaller than the 1 m minimum distance between the vehicles considered in (\textit{A1}). 
	\vspace{2mm}
	\item \textit{A4:} Transmissions by the VLC units do not interfere. This can be achieved via special medium access control mechanisms, e.g., by assigning each unit to a separate transmission frequency band \cite{tx_interference_ref}.
	\vspace{3mm}
\end{itemize}

\noindent We call the vehicle that is being positioned the \quotes{target}, and the vehicle that is estimating the position the \quotes{ego}. {\color{black}The two vehicles do not need to be identical}. Fig. \ref{sysmdl} depicts this system model for the case of a target vehicle (green) being followed by an ego vehicle (red), {\color{black}but both vehicles can take either role.}

\subsection{Vehicular VLP Problem - Technical Definition}

Vehicular VLP methods use RX VLC signals to estimate TX relative positions, which are denoted as $(x_{ij},~y_{ij})$ in this paper with regards to the indices $i, j \in \{1,2,3,4\}$ for vehicular VLC RXs and TXs, respectively. For determining the exact 2D relative location of the target vehicle, positioning two TX units on the target vehicle with respect to two RX units on the ego vehicle is necessary, as discussed in \cite{soner_tvt}. However, since the two ego RX units are simply separated by $L$ in the lateral axis of the ego frame of reference, estimating positions with respect to only one RX is sufficient, e.g., only $(x_{11}, y_{11})$ and $(x_{12}, y_{12})$ for RX 1 in Fig. \ref{sysmdl}. For this reason, the $i$ subscript in $(x_{ij},~y_{ij})$ can be dropped when describing TX position coordinates, i.e., $(x_{j},~y_{j}) \! = \! (x_{1j},~y_{1j})$ for TX $j$, $j \in \{1,2\}$, as in Fig. \ref{sysmdl}. Additionally, if vehicles are longitudinally parallel, the following holds: $(x_{2}, y_{2}) = (x_{1} + L, ~y_{1})$, and estimating only one position, $(x_{1}, y_{1})$, is sufficient. {\color{black}Note that RX separation $L$ is naturally known by the ego vehicle, but it does not necessarily have to be equal to TX separation, and TX separation is assumed to be unknown.}

Accordingly, although estimating position for the two TX units is necessary for completely characterizing localization performance, analyses in this paper show results and derivations for only one TX due to the following reasons: \linebreak1) Extending the analyses to a second TX is straightforward, 2) the position estimation error for one TX does not differ significantly from that of the other due to the physical proximity between the two, and 3) this allows fair comparison with methods that assume longitudinally parallel vehicles (e.g., \cite{roberts_pdoa}) which converge to the closest solution that satisfies the parallel assumption even when the actual orientations of vehicles do not. Therefore, estimation error is defined as

\vspace{-1mm}
\begin{equation}
	\label{accuracy}
	e = ~\sqrt{ \left( x_{1} - \widehat{x_{1}}  \right)^2 +  \left( y_{1} - \widehat{y_{1}}  \right)^2}~,
\end{equation}
\vspace{-3mm}

\noindent where $(\widehat{x_{1}},~\widehat{y_{1}})$ is the position estimation for TX 1 (~$\widehat{}$~ denotes measured or estimated quantities in the rest of the text), and $e$ is the associated error for a given estimate. Note that due to high mobility and finite estimation rate, $e$ varies within an estimation period. The maximum such $e$ value is considered as the error for that estimation period and faster relative vehicle movement or lower estimation rate induces higher error. This mobility-induced error forms the basis of the 50 Hz minimum estimation rate requirement \cite[Fig. 2]{soner_tvt}. The other main source of error is the random VLC channel noise which necessitates evaluating a distribution of $e$ values sampled over the noise process rather than individual $e$ samples. Accordingly, the term \quotes{cm-level accuracy} in this paper refers to the distribution of $e$ having one standard deviation above the mean smaller than 10 cm \footnote{Note that in the zero-mean case (i.e., unbiased estimation), this corresponds directly to standard deviation being less than 10 cm.}.

\subsection{Mathematical Model of Received VLC Signals}

The following received VLC signal model is considered:

\vspace{1mm}
\begin{equation}
	\label{rx_model_mag}
	r_{i}(t) = \sum\limits_{j} r_{ij}(t)  + \mu_{i}(t) ~~,~~ r_{ij}(t) = H_{ij}\cdot s_{j}(t - \tau_{ij})~,
\end{equation}
\vspace{1mm}

\noindent where $t$ is time, $r_{i}$ is the total received photocurrent signal at RX $i$, $s_{j}$ is the transmitted photocurrent signal at TX $j$, $r_{ij}$ is the contribution of $s_{j}$ to $r_{i}$, $\mu_{i}$ is the photocurrent noise, and $H_{ij}$ and $\tau_{ij}$ are respectively the geometric channel gain and the finite propagation time from TX $j$ to RX $i$. Since TX signals are directional, only two TXs on a given face of the target vehicle contribute to Eqn. (\ref{rx_model_mag}) for a given RX on the ego vehicle, i.e., $j$ is either $\in \{1,2\}$ or $\in \{3,4\}$ for a given $i \in \{1,2,3,4\}$ depending on vehicle orientation, and the two TX components arriving at RX $i$ can be processed separately as $r_{ij}$ in Eqn. (\ref{rx_model_mag}) as per assumption \textit{A4}. 

Expressions for $H_{ij}$, $\tau_{ij}$ and $\mu_{i}$ rely on the point-source approximation\footnote{{\color{black}Note that this does not imply a Lambertian radiation pattern. It only implies that the source is not extended, thus, the far-field pattern is considered.}} of TXs from the perspective of RXs, which simplifies expressions for $H_{ij}$ and $\tau_{ij}$, and indirectly, also $\mu_{i}$. Accordingly, the channel gain $H_{ij}$ can be expressed as


\vspace{1mm}
\begin{equation}
	\label{rx_model_gain}
	H_{ij} =  \gamma_{i} \rho_{i}(\theta_{ij}) \iint \limits_{S_{ij}} \gamma_{j} \rho_{j}(S)~ dS ~,~ S_{ij} \propto \frac{A_i \cos(\theta_{ij})}{d_{ij}}~,
\end{equation}
\vspace{1mm}


\noindent where $d_{ij}$ and $\theta_{ij}$ are the relative distance (i.e., range) and AoA (i.e., bearing) of TX $j$ relative to RX $i$, respectively; $\rho_{j}$ and $\rho_{i}$ are the normalized positive-definite TX beam pattern and RX ``reception'' pattern, respectively; $\gamma_{j}$ and $\gamma_{i}$ are the TX electrical-to-optical gain and RX optical-to-electrical, i.e., photodiode sensitivity respectively; and $S_{ij}$ is the solid angle subtended by the active area of the RX $i$ ($\triangleq \! A_i$) with respect to TX $j$ \cite{solid_angle}. 

Lambertian source models turn Eqn. (\ref{rx_model_gain}) into a closed form expression. However, since non-Lambertian beam patterns can also be used, we provide the more general Eqn. (\ref{rx_model_gain}), which can easily be converted to an approximate Lambertian model when necessary, as done so in \cite{heidi_singleStep, pw_lambert, becha_ranging}. $\theta_{ij}$ and $d_{ij}$ in Eqn. (\ref{rx_model_gain}), and $\tau_{ij}$ in Eqn. (\ref{rx_model_mag}) are expressed as

\vspace{1mm}
\begin{equation}
	\label{rx_model_phs}
	\small
	d_{ij} \!=\! \sqrt{{x_{ij}}^2 \! + \! {y_{ij}}^2}~~,~~\theta_{ij} = \arctan \left( \frac{x_{ij}}{y_{ij}} \right)~~,~~\tau_{ij} = \frac{d_{ij}}{c}~,
\end{equation}
\vspace{1mm}

\noindent where $c$ is the speed of light. 

$\mu_{i}$ is composed of shot noise on the receiving p-i-n photodetector (PD) and thermal noise on the FET-based front-end transimpedance amplifier (TIA) that follows the PD \cite{smith1980receiver}. The combined noise is zero mean additive white Gaussian (AWGN) with variance \cite[Eqn. (18)]{kahn}:

\begin{subequations}
	\vspace{1mm}
	\begin{equation}
		\label{rx_noise_total}
		\sigma_{\mu_{i}}^2 = \sigma_{shot_{i}}^2 + \sigma_{thermal_{i}}^2~,
	\end{equation}
	\vspace{1mm}
	\begin{equation}
		\label{rx_noise_shot}
		\sigma_{shot_{i}}^2 = 2 q \gamma_{i} P_{i} B_{i} + 2 q I_{bg,i} I_{B2} B_{i}  ~,
	\end{equation}
	\vspace{1mm}
	\begin{equation}
		\label{rx_noise_thermal}
		\sigma_{thermal_{i}}^2 = 4k T_{i} \left( \frac{1}{R_{i}} I_{B2} B_{i} \! + \!  \frac{\left(2 \pi C_{i} \right)^2}{g_{i}} \Gamma I_{B3} B_{i}^3 \right)~,
	\end{equation}
	\vspace{1mm}
\end{subequations}

\noindent where $q$ is the Coulomb electron charge, $k$ is the Boltzmann constant, $P_{i}$ is the optical signal power on RX $i$, $I_{bg,i}$ is the background illumination current, $B_{i}$ is the front-end bandwidth, $T_{i}$ is the circuit temperature, $R_{i}$ is the front-end resistance (i.e., TIA feedback gain term), $C_{i}$ is the input capacitance due to the photodiode and the FET, $g_{i}$ is the FET transconductance, and $\Gamma$, $I_{B2}$ and $I_{B3}$ are unitless factors for FET channel noise and noise bandwidth determined by the signal shape \cite{kahn}. For an optimal TIA (i.e., proper loop compensation and impedance matching such that bitrate is equal to $B_{i}$ \cite{smith1980receiver}), Eqn. (\ref{rx_noise_thermal}) is typically reorganized using $R_{i} = G/(2 \pi B_{i} C_{i})$ where $G$ is the ``open-loop voltage gain'', and the front-end circuit gain is independent of transistor parameters, i.e., $R_{i}$ determines the transimpedance gain which turns the received photocurrent $r_{i}$ into a voltage signal.

We ignore the following minor effects: Popcorn noise due to silicon defects are absent in modern components. Flicker, i.e., $1/f$ noise, is also ignored since VLC operation is not near DC. Furthermore, random fluctuations on $H_{ij}$ and $\tau_{ij}$ due to atmospheric turbulence are ignored since LEDs are non-coherent. Similarly, Doppler effects, which would make $\tau_{ij}$ time-dependent, are also ignored since they are shown to have a negligible effect on positioning performance \cite{becha_vlcr_experimental}. 


\section{Vehicular VLP Methods}

This section first reviews parameter measurement techniques and existing algorithms that use those techniques for position estimation in vehicular VLP, and then {\color{black}proposes a novel algorithm that combines bearing and range measurements for positioning.}

\vspace{-1mm}
\subsection{Parameter Measurement from Received VLC Signals}

Vehicular VLP methods utilize parameter measurement techniques that use received VLC signal samples to produce bearing/range measurements. Specifically, these techniques make observations on $H_{ij}$ or $\tau_{ij}$, which are then translated to direct relative TX bearing ($\theta_{ij}$) or range ($d_{ij}$) measurements to one RX, or their differential versions between two RXs when direct measurements are not available. Differential range and bearing are respectively defined as 

\vspace{1mm}
\begin{equation}
	\label{diffs}
	\Delta d_{il/j} \!=\! d_{ij} \!- \! d_{lj}~~~~ \text{and}~~~~ \Delta \theta_{il/j}\! =\! \theta_{ij}\! -\! \theta_{lj}~~,
\end{equation}
\vspace{0mm}

\noindent where $l \in \{1,2,3,4\}, ~l \! \neq \! i$, denotes the index for the second RX unit on the same face as RX $i$. This type of signal-based physical parameter measurement has long been an active area of research due to its useful applications in communication (channel estimation \cite{acoustic_channelest, scoleri_channelest, vlc_channelest}) as well as radio-navigation (i.e., RADAR). While the main challenges in the area arising from fading and other multi-path effects are highly pronounced when RF signals are used, vehicular VLC signals are practically devoid of these effects since intensity modulation and direct detection (IM/DD \cite{vlc_survey}) is used and the channel is LoS-dominant for the distances relevant in collision avoidance and platooning \cite{vlc_channel_wnl}, enabling high-accuracy measurement \cite{mfkeskin_ranging, mfkeskin_proceeding, wang_crb_toa, heidi_crb_aoa}. Techniques used for bearing and range measurement in vehicular VLP, categorized by their use of $H_{ij}$ or $\tau_{ij}$, are reviewed in the following. 

\vspace{1mm}
\subsubsection{Techniques Using Propagation Delay Observations ($\tau_{ij}$)}

Observations on propagation delay are a perfect fit for range measurements since TX-RX distance is directly proportional to $\tau_{ij}$ as per Eqn. (\ref{rx_model_phs}). In VLC systems employing multi-carrier modulation strategies like orthogonal frequency division multiplexing (OFDM), observations on $\tau_{ij}$ can be made via measuring phase difference over individual carriers or pilot symbols \cite{ofdm_channelest_vlc, ofdm_channelest, ofdm_timedelay}. Without loss of generality, we assume a constantly available simple sinusoidal symbol or pilot (i.e., pure tone) for simplifying the analyses in this paper, reducing the $\tau_{ij}$ computation problem to the estimation of the phase shift between that sinusoid and its propagation-delayed version. Numerous techniques are available for this computation \cite{dsp_phaseshift}, and some of them have been utilized in vehicular VLP literature for range measurements \cite{autodigital1, becha_ranging, roberts_pdoa}. Note that the matched-filter approach is not feasible for vehicular VLP since cm-level accuracy would require \textgreater10 GHz sampling rates, increasing system cost beyond practical levels.

For direct range measurements, \cite{becha_ranging} presents an auto-digital based technique \cite{autodigital1}: VLC units on the ego and target vehicles exchange a high frequency square wave back and forth, and the ego vehicle measures the phase shift between the initial TX signal and the round-trip-propagated RX signal. The channel-corrupted RX signal on the ego vehicle first gets converted to a logic signal by a high-speed comparator \cite{becha_vlcr_experimental} (i.e., zero-crossing detection) and then gets heterodyned to a lower frequency along with the original TX signal. Since heterodyning increases phase resolution considerably, the shift can be measured in high precision by sampling the logical XOR of the two signals with a high-frequency clock and counting the number of high pulses. While this means the quantized measurement resolution is fundamentally limited by the clock frequency and the heterodyning factor, the quantization intervals can be made small with high-frequency hardware components, e.g., \textless1 cm for a 100 MHz clock. Although this technique requires high-frequency digital circuits, it has very low complexity and can be completely integrated into a CMOS pipeline since logic (i.e., virtually 1-bit) signals are used. Moreover, sine waves of the same frequency can be transmitted over the channel instead of square waves when narrow band-pass filters are utilized since comparators already convert them to logic signals on the RX end. This makes the technique realizable with current low-bandwidth high-power automotive LEDs and multi-carrier wide-band VLC modulation strategies.

\begin{figure}[t]
	\centering
	\includegraphics[width=0.49\textwidth]{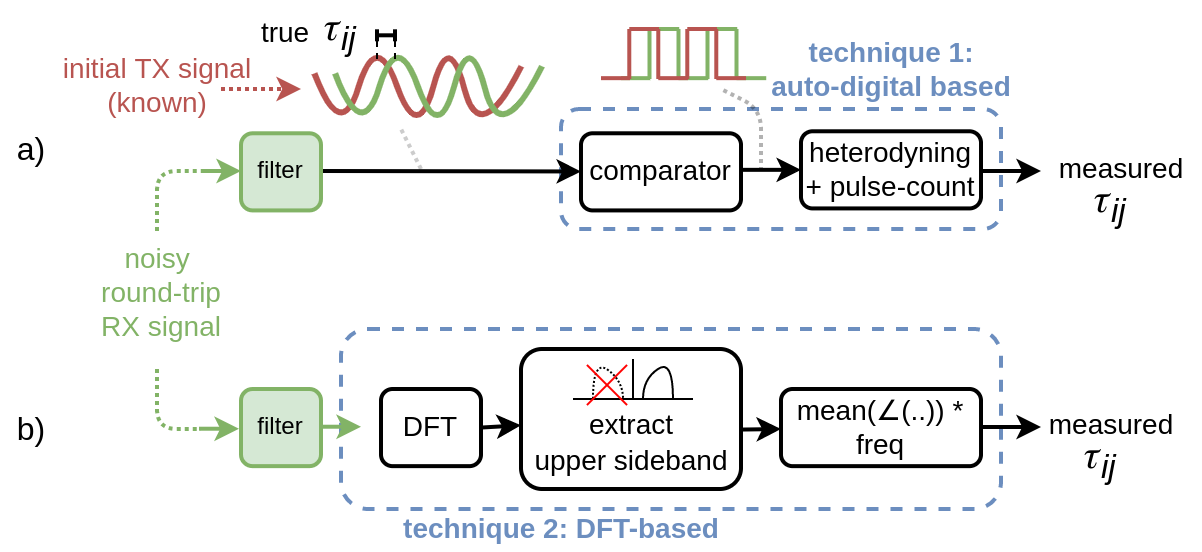}
	\vspace{-3mm}
	\caption{(a) Auto-digital based and (b) DFT-based direct range measurement techniques using observations on $\tau_{ij}$ and Eqn. (\ref{rx_model_phs}). Technique (b) is originally proposed for measuring differential range over 2 RX signals.}
	\vspace{3mm}
	\label{rangemeas}
\end{figure}

For differential range measurements to two RXs (RX $i$ and RX $l$) from one TX, \cite{roberts_pdoa} presents a DFT-based technique: The upper DFT sideband of one RX signal (RX $i$) is multiplied with the complex conjugate of its counterpart for the other RX signal (RX $l$). The mean of the phase for the result is an estimate of the phase shift between the two RX signals, for the same TX. Although previously unexplored in vehicular VLP, this technique can also be used for direct range measurement when the scenario presented in \cite{becha_ranging} is used (i.e., RX $i$ and RX $l$ are replaced by a known TX signal and its round-trip-propagated RX version). These two measurement techniques are depicted on the same block diagram in Fig. \ref{rangemeas}. While the complexity of the DFT-based technique \cite{roberts_pdoa} is much higher than the auto-digital technique \cite{becha_ranging} due to DFT use and the  analog-to-digital conversion requirement (ADC, much higher than 1-bit precision), it typically provides higher accuracy against noise due to averaging as will be shown via simulations in Section IV. Similarly, the auto-digital measurement principle \cite{becha_ranging} can be applied for differential range measurement, but since differential values are very small, using this technique would be impractical due to the extremely high clock speed required for that level of precision.

Lastly, $\tau_{ij}$ observations do not provide means for bearing measurement as evident from Eqn. (\ref{rx_model_gain}). While derivative measurements are possible, e.g., differential bearing measurement using direct range to two RXs, $L$ and the law of cosines, this simply transfers the uncertainty from one measurement type to the other, thus, is not a dedicated technique.

\vspace{1mm}
\subsubsection{Techniques Using Channel Gain Observations ($H_{ij}$)}

The relationship between $H_{ij}$ and the relative TX-RX distance is straightforward for the LoS vehicular VLC channel: $H_{ij}$ is inversely proportional to the square of the propagation distance \cite{vlc_channel_wnl}. Accordingly, for sinusoidal signals, an accurate observation on $H_{ij}$ can be made via computing received signal strength (RSS) by integrating the square of $r_{ij}(t)$ over an interval that captures many $s_j(t)$ cycles \cite{heidi_crb_aoa}. However, extracting range information from that RSS info requires exact knowledge of instantaneous TX power (cf. Eqn. (\ref{rx_model_gain}) and \cite[Eqn. (4)]{mfkeskin_ranging}). This is highly unlikely even after a potential future standardization of vehicular VLC TX power since the electrical reference on a vehicle (i.e., a regulated version of the battery voltage) is subject to broadband random fluctuations (both in combustion \cite{ice_voltage_fluctuations} and electrical vehicles \cite{elecveh_voltage_fluctuations}), making exact instantaneous knowledge of TX power infeasible. Therefore, RSS-based direct ranging is not feasible for use in vehicular VLP. 

RSS-based differential range measurement is also not feasible in vehicular VLP for a similar reason: In order to extract $\Delta d_{il/j}$ from RSS values, the technique would require exact knowledge of the TX radiation pattern ($\rho_{j}$) and of the instantaneous solid angles subtended by each RX ($i$ and $l$) to the TX. Since this is not realistic, RSS-based differential range measurement is also not feasible for use in vehicular VLP. 

\begin{figure}[t]
	\centering
	\includegraphics[width=0.44\textwidth]{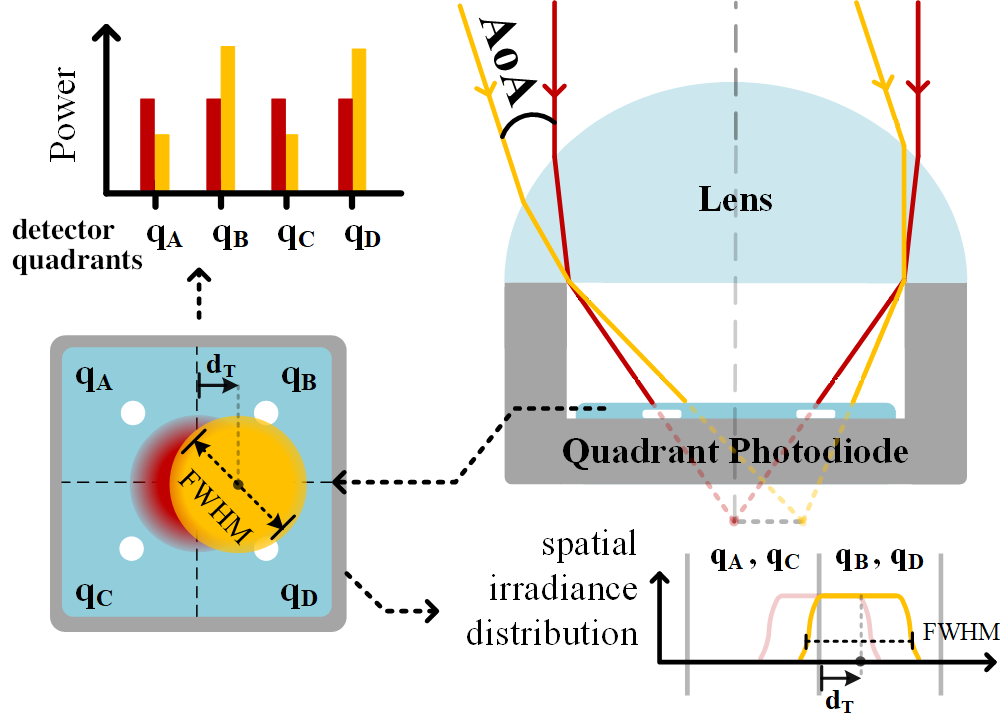}
	\vspace{1mm}
	\caption{Direct bearing measurement using observations on $H_{ij}$ (\quotes{Power}) and Eqn. (\ref{rx_model_gain}). Image adapted from \cite{soner_tvt}. Other angular diversity RX designs with, e.g., photodiode arrays of different shapes and number of elements, and/or different front-end optics (e.g., aperture-only) \cite{ada_receivers}, can also be utilized.}
	\vspace{-1mm}
	\label{bearingmeas}
\end{figure}

RSS can still be used for bearing measurements via angular diversity receivers: An RX that employs closely-packed detectors with different angular \quotes{reception} patterns (i.e., $\rho_{i}$) has extremely close AoA values and different RSS values at all detectors, enabling precise AoA measurement. An example that utilizes a quadrant detector \cite{soner_tvt} is depicted in Fig. \ref{bearingmeas}. Accuracy of these measurements fundamentally depend on how close the detectors are\footnote{Distance between the detectors should be much smaller than their distances to the TX to keep both the bearing and the $S_{ij}$ the same for all detectors.} and how different their $\rho_i$ patterns are\footnote{RSS differences only depend on $\rho_{i}$ patterns because small-scale fading is insignificant due to detector sizes being millions of wavelengths \cite[Fig. 2a]{kahn}}. While there have been many such RX designs in indoor multiple-input-multiple-output (MIMO) VLC \cite{ada_receivers, wang_hemisphericalAnalyz, wang2014prism, sphericalRX} and associated AoA measurement techniques for indoor VLP \cite{esteve_qadaish, qadaplus, cornercuberx, steendam_sqrarray, qada_origins, qada_esteve}, only a few of these design ideas are feasible in the vehicular domain due to high mobility and harsh road conditions. Specifically, early works have considered pyramidal detectors \cite{ir_bizimgeo}, but such tilted-detector based techniques, which have also been studied in indoor cases \cite{optimal_tilted_aoa, tilted_aoa_circular, tilted_heidili, tilted_unlu}, suffer in precision and field-of-view (FoV). More recently, a planar quadrant-photodiode based RX with front-end optics (named \quotes{QRX}) has been shown to achieve high-precision and high-FoV direct bearing measurement for vehicular localization in \cite{soner_tvt}.

RSS-based differential bearing measurement techniques are still unexplored in vehicular VLP. Therefore, to enable performance evaluations for the differential bearing based VLP methods later in the paper, subtraction of two direct bearing measurements will be considered as a differential bearing measurement even though it is a derivative measurement. 

A summary of all individual parameter measurement techniques that have been described so far with respect to their use in vehicular VLP, are provided in Table \ref{param_meas_table}.

\begin{table*}[!t]
	\renewcommand{\arraystretch}{1.6}
	\caption{Parameter Measurement Techniques Used in Vehicular VLP}
	\label{param_meas_table}
	\centering
	\begin{tabularx}{\textwidth}{r|Y|Y|Y|Y}
		\hline
		measurement type & direct range ~~~~~~~~~~~~$d_{ij}, ~d_{lj}$ &  differential range ~~~~~~~~~$\Delta d_{il/j}$ & direct bearing ~~~~~~~~~$\theta_{ij},~\theta_{lj}$ &  differential bearing ~~~~~~~~$\Delta \theta_{il/j}$ \\
		\hline
		\hline
		using $\tau_{ij}$ observations & auto-digital \cite{becha_ranging}, ~~~~~(DFT-based) \textsuperscript{a} \cite{roberts_pdoa} & DFT-based \cite{roberts_pdoa}&  \multicolumn{2}{c|}{no direct relation, cf. Eqn. (\ref{rx_model_gain})} \\
		\hline
		using $H_{ij}$ observations & \multicolumn{2}{c|}{not feasible \textsuperscript{b}} & \quotes{QRX} \cite{soner_tvt} & \textendash ~\textsuperscript{c} \\
		\hline
	\end{tabularx}
	\vspace{2mm}
	\justify \footnotesize \textsuperscript{a} This was previously unexplored. We use it to realize the novel algorithm described in Section III-C and demonstrate its performance in Section IV.
	\vspace{-2mm}
	\justify \footnotesize \textsuperscript{b} These require knowledge of exact TX power or exact solid angle subtended by RXs on TX $\rho_j$, both of which are unfeasible in vehicular use cases.
	\vspace{-2mm}
	\justify \footnotesize \textsuperscript{c} No techniques have been proposed so far. We consider subtraction of two direct bearing measurements for this type to facilitate later related analyses.
	\vspace{-1mm}
\end{table*}


\subsection{Existing Positioning Algorithms in Vehicular VLP}

Positioning algorithms in vehicular VLP use the available parameter measurements described in Section III-A and combine them to estimate position. Based on system model assumption (\textit{A2}), at most two RXs on a given face of the ego vehicle are considered available in this paper for positioning. Positioning algorithms that use this minimal configuration can be categorized with respect to their use of the following:

\begin{itemize}
	\vspace{1mm}
	\item differential (as in Eqn.(\ref{diffs})) vs. direct measurements,
	\vspace{1mm}
	\item for bearing vs. range parameters, and
	\vspace{1mm}
	\item \quotes{classical} vs. \quotes{running} position fixing.
	\vspace{1mm}
\end{itemize}

\begin{figure}[t]
	\centering
	\includegraphics[width=0.48\textwidth]{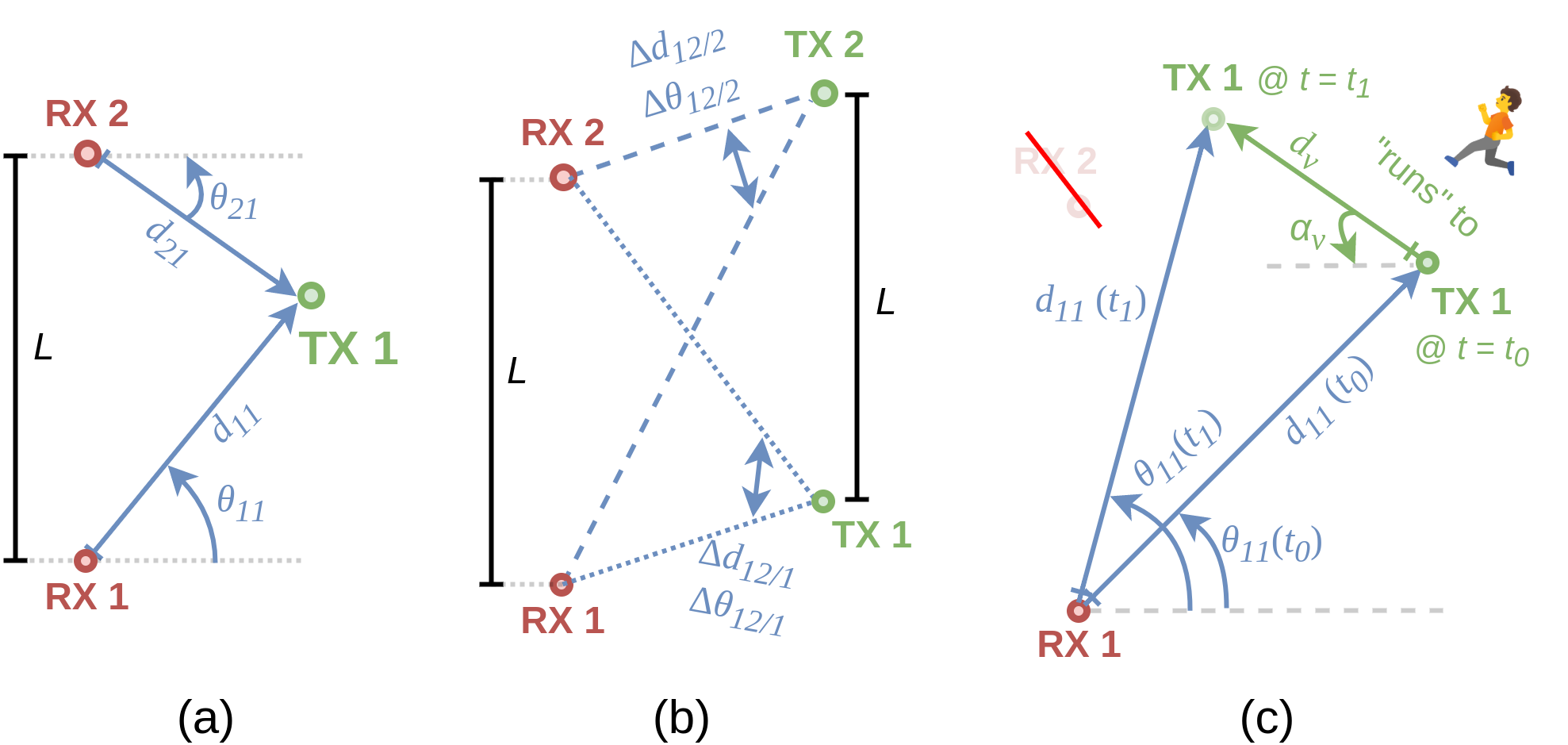}
	\vspace{2mm}
	\caption{(a) 2 RXs and 1 TX are required for a classical fix with direct measurements for bearing ($\theta_{ij}$) or range ($d_{ij}$, where $i$ and $j$ are indices for RX and TX units, respectively). (b) 2 RXs and 2 TXs that are longitudinally parallel are required for a classical fix with differential measurements for bearing ($\Delta\theta_{il/j}$) or range ($\Delta d_{il/j}$ as defined in Eqn. (\ref{diffs}). (c) A running fix with direct measurements only needs 1 RX and 1 TX, but it also needs extra information (i.e., relative target heading, $\alpha_v$, and travel distance, $d_v$) to estimate position for two time instants (i.e., $t=t_0$ and $t=t_1$) \cite{soner_pimrc}.}
	\label{fixing}
	\vspace{-2mm}
\end{figure}

\quotes{Classical} position fixing considers using triangulation or trilateration, i.e., a TX gets within the LoS of the two RXs, and the measurements from the two RXs, together with the known {\color{black}ego vehicle light separation} $L$ in between {\color{black}(target vehicle light separation is not known)}, form a determined set of equations. On the other hand, \quotes{running} position fixing can be used {\color{black}as a compromise when one of the TXs leaves the FoV of one of the ego RXs}, which frequently happens due to high mobility. The running fix, heavily used in marine and aviation disciplines, uses two consecutive measurements from only one RX to a moving TX and assumes that the TX relative heading ($\alpha_v$) and distance traveled ($d_v$) within that time is known {\color{black}(can be transmitted over the VLC link \cite{soner_pimrc})}. Classical fixing for direct and differential bearing/range measurements, and running fixing with direct range measurements, are pictured in Fig. \ref{fixing}a, Fig. \ref{fixing}b and Fig. \ref{fixing}c, respectively. {\color{black}While measurements from more than two RXs and/or more measurements in time could potentially improve performance (e.g., via least squares), only the minimal configuration is considered here to assess the worst case performance (cf. \textit{A2}).}

{\color{black}Feasible combinations of the bearing-only or range-only algorithms depicted in Fig. \ref{fixing} have been proposed in the literature up to now, as summarized in Table \ref{algorithms}. Classical position fixing with direct measurements, i.e., triangulation \cite{triangulation_bizimki} has been explored for bearing measurements as in \cite{soner_tvt}, and trilateration has been explored for range measurements as in \cite{becha_positioning}. Considering differential measurements instead of direct measurements, this has been explored for ranges by \cite{roberts_pdoa} and for bearings by \cite{techreport_vlp_sonercoleri}. When the target TX is not in sight of one of the two RXs, a running position fix can be made with 1 RX and 1 TX as a compromise. This has been explored for direct ranges by \cite{techreport_vlp_sonercoleri} and for bearings by \cite{soner_pimrc}.} Differential measurements cannot be used for running fixes because that would require the target vehicle to move only sideways, conflicting with Ackermann steering rules (i.e., no side-slip \cite{ackermann}). {\color{black}Hybrid methods that combine bearing and range were previously unexplored since an RX design that can simultaneously measure both parameters did not exist. A novel method that does this is proposed in Section III-C. The theoretical performance of all methods can be analyzed by evaluating the CRLBs that are derived in \cite{techreport_vlp_sonercoleri} under realistic driving scenarios. These evaluations are presented alongside simulation results in Section IV to assess their performance.}

\begin{table}[b]
	\renewcommand{\arraystretch}{1.6}
	\caption{{\color{black}Vehicular VLP Algorithms Using a Single Parameter}}
	\label{algorithms}
	\centering
	\begin{tabu}{p{0.1\linewidth}|p{0.15\linewidth} | p{0.15\linewidth} | p{0.15\linewidth} | p{0.15\linewidth}|}
		\cline{2-5}
		& \multicolumn{2}{c|}{direct meas.}& \multicolumn{2}{c|}{differential meas.}\\
		\cline{2-5}
		& classical fix &  running fix & classical fix &  running fix \\
		\hline
		range & \checkmark, \cite{becha_positioning} & {\color{black}\checkmark, \cite{techreport_vlp_sonercoleri}} & \checkmark, \cite{roberts_pdoa} & $\times$ \textsuperscript{a} \\
		\hline
		bearing & \checkmark, \cite{soner_tvt} & \checkmark, \cite{soner_pimrc} & {\color{black}\checkmark, \cite{techreport_vlp_sonercoleri}} & $\times$ \textsuperscript{a}  \\
		\hline
	\end{tabu}
	\vspace{2mm}
	\justify \footnotesize \textsuperscript{a} not feasible since it strictly requires the vehicle to move only sideways.
	\vspace{-1mm}
\end{table}

\subsection{{\color{black}A Novel Hybrid Vehicular VLP Method}}

{\color{black}Current state-of-the-art classical position fixing methods utilize either direct bearing measurements as in \cite{soner_tvt} or direct range measurements as in \cite{becha_positioning}, but not both. The analyses in \cite{techreport_vlp_sonercoleri} make it evident that combining these two methods would improve performance, i.e., range-based positioning provides higher accuracy in the longitudinal axis, and vice versa for bearing and the lateral axis. The main challenge is designing an RX that allows for measuring both simultaneously. We propose a solution: The QRX, which is originally designed for bearing measurements in \cite{soner_tvt}, can be repurposed to simultaneously do direct range measurement with the DFT-based method in \cite{roberts_pdoa} (auto-digital \cite{becha_ranging} would require additional CMOS circuitry). The TX signal, which already gets decoded by the QRX VLC subsystem for bearing measurement, can be used in place of the second RX signal to convert the DFT-based differential method to a direct one as described in Section III-A-1. 

One consideration for doing range measurements with such an imaging receiver would be the refraction-induced time-of-flight delay on the signal inside the lens. However, since the lens-sensor distance on the QRX is only a few millimeters, this delay is negligible compared to the TX-RX distance, which is on the order of meters. Therefore, the QRX can be used for range measurements as well bearing measurements simultaneously, and realize the envisioned hybrid vehicular VLP method.}

\section{Simulations}

{\color{black}The simulations in this section have three goals: 1) benchmarking the performance of all existing vehicular VLP methods under realistic collision avoidance and platooning scenarios to identify the best performers, 2) characterizing the best performers against severe VLC channel noise, adverse weather conditions, high mobility, real measured (incl. asymmetric) light patterns and different vehicle dimensions, and 3) validating the theoretical performance of these algorithms by evaluating their respective Cramer-Rao Lower Bounds (CRLB) for positioning accuracy (derived in \cite{techreport_vlp_sonercoleri}) alongside simulation results.} A custom simulator is built for this purpose in Python language and all related source material (i.e., code and related documentation) is made available on GitHub \cite{github_vvlp}. 

Simulator setup parameters are given in Table \ref{table_sim}. In all simulations, a leading target vehicle transmits VLC signals from its tail light towards the ego vehicle (Fig. \ref{sysmdl}); this is the worst case scenario since taillights have lower optical power (2 W) than headlights. {\color{black}The taillight is considered to be a Lambertian pattern of 20\textdegree ~half-power angle (order m=11) throughout most simulations since it is the closest analytical pattern to taillight regulations \cite{becha_ranging} as well as measured patterns \cite{turan_3tx}, but we also evaluate different patterns such as wider Lambertians \cite{pw_lambert} and real asymmetric patterns \cite{memedi_headlightpatterns} to quantify their effect on performance.} Pure tone modulation at 1 MHz \cite{becha_ranging} is considered since it is a realistic pass-band limit for current high-power automotive LEDs. The signal is band-pass filtered at 100 kHz bandwidth around the 1 MHz carrier on the RX end as in \cite{becha_vlcr_experimental}. To facilitate fair comparison and practical relevance, all methods are configured for 100 Hz positioning rate (i.e., the rate requirement is satisfied, and simulations test the accuracy requirement), ADCs have a moderate 10 MSPS conversion rate, and all methods utilize the \quotes{QRX} unit \cite{soner_tvt} {\color{black}following the discussion in Section III-C}. To further improve practical relevance, we choose {\color{black}the solid model of} a very simple plano-convex lens from the Edmund Optics catalog (\#67-149) during QRX design, and accordingly tune for $\approx\pm60^\circ$ FoV using ray-optics simulations \footnote{\href{https://github.com/mjhoptics/ray-optics}{https://github.com/mjhoptics/ray-optics}}. A wide range of realistic ambient light induced noise conditions (night, day, direct sunlight \cite{moreira1997optical}) and weather-induced attenuation conditions (clear, foggy, rainy \cite{fograin}) are considered in the simulations. 

Accordingly, we present seven evaluation scenarios (\textit{ES\#}): 

\begin{table}[!t]
	\renewcommand{\arraystretch}{1.6}
	\caption{Simulator Setup Parameters}
	\label{table_sim}
	\centering
	\begin{tabular}{c|l|l}\hline
		\multirow{6}{*}{TX} & Signal & pure tone, 1 MHz \\
		\cline{2-3}
		& Power &  $\gamma_j \cdot \max(\lvert s_j \lvert)$ = 2 W (tail light) \\
		\cline{2-3}
		& {\color{black}Patterns} & {\color{black}various Lambertians \cite{becha_ranging, pw_lambert} and real \cite{memedi_headlightpatterns}} \\
		\cline{2-3}
		& \multirow{3}{*}{Attenuation} & clear: - \\ 
		\cline{3-3}
		& & heavy rain ($\approx$10 mm/hr): 0.1 dB/m \cite{fograin}\\
		\cline{3-3}
		& & dense fog ($\approx$200 m):  0.3 dB/m \cite{fograin}\\
		\hline
		\hline
		\multirow{9}{*}{QRX} & $\gamma_i$, $g_i$, $A_i$ \textsuperscript{a} & 0.5 A/W, 30 mS, 31.2 mm\textsuperscript{2} \\
		\cline{2-3}
		& $B_i$, $C_i$, $R_i$ & 10 MHz \textsuperscript{b}, 45 pF, 2.84 k$\Omega$, i.e., $G\approx10$ \\
		\cline{2-3}
		& Factors  & $\Gamma$=1.5 , $I_{B2}$=0.562 , $I_{B3}$=0.0868 \\
		\cline{2-3}
		& Temperature & $T$=298 K \\
		\cline{2-3}
		& \multirow{2}{*}{$I_{bg,i}$~\cite{moreira1997optical}} & night-time: 10 $\mu$A \\ 
		\cline{3-3}
		& & day, indirect sun: 750 $\mu$A\\
		\cline{2-3}
		& Lens & N-BK7 (n=1.52), diameter = 9.0 mm\\
		\cline{2-3}
		& Detector  & $d_{H}$ = 6.3 mm , $d_{X}$ = 1.9 mm, see \cite{soner_tvt} \\
		\cline{2-3}
		& FoV & $\pm60$\textdegree, see \cite{github_vvlp} for ray-optics simulations \\
		\cline{2-3}
		\hline
		\hline
		\multirow{4}{*}{Vehicle} & Length & 5 m  \\
		\cline{2-3}
		& {\color{black}Light Height} & {\color{black}target = [1.1m : 1.7m], ego = 1.1 m} \\
		\cline{2-3}
		& {\color{black}Width ($L$)} & {\color{black}target = 1.5 m, ego = 1.6 m} \\
		\cline{2-3}
		& Steering & Ackermann \cite{ackermann} (small sideslip) \\
		\hline		
	\end{tabular}
	\justify \footnotesize \textsuperscript{a} This is the area of the aperture stop, which is 6.3 mm in diameter.
	\vspace{-2mm}
	\justify \footnotesize \textsuperscript{b} This is ADC output, but effectively, $B_i =$ 100 kHz due to filtering \cite{becha_vlcr_experimental}.
	\vspace{-2mm}
\end{table}

\vspace{2mm}
\begin{itemize}
	\item \textit{ES1 - Comparing Range Measurement Techniques}: We simulate the two range measurement techniques used in vehicular VLP, i.e., auto-digital \cite{becha_ranging} and DFT-based \cite{roberts_pdoa}, for static target vehicle locations on a 1-20 m straight test-track (i.e., bearing$=$0) to characterize and compare their performance.
	\vspace{3mm}
	\item {\color{black}\textit{ES2 - Benchmark, All Methods, Lane Change Scenario}: We simulate all six existing vehicular VLP methods listed in Table \ref{algorithms} as well as the novel hybrid algorithm proposed in Section III-C in a lane change scenario to benchmark their performance in a realistic driving setting. The target vehicle passes over the lane of the ego vehicle from the left, moderate speed, as described in \cite{techreport_vlp_sonercoleri}.}
	\vspace{3mm}
	\item {\color{black}\textit{ES3 - Validating Theoretical Performance Analyses}}: We evaluate the CRLBs of all classical position fixing methods at various static target vehicle locations over the operational range to validate the theoretical performance analyses in \cite{techreport_vlp_sonercoleri}. The vehicles stay longitudinally parallel for fair comparison between methods utilizing direct and differential measurements.
	\vspace{3mm}
	\item {\color{black}\textit{ES4 - Characterizing Vehicular VLP Operational Range}: We evaluate the performance of the top three performers from \textit{ES2} over a grid that covers a typical 3-lane road (a 6 m x 14 m zone in front of the ego vehicle) to determine the operational range of the best-performing vehicular VLP methods.}
	\vspace{3mm}
	\item {\color{black}\textit{ES5 - Weather Conditions, Collision Avoidance Scenario}: We use a recorded dynamic collision threat scenario extracted from the INTERACTION dataset \cite{interaction_dataset} to characterize the performance of the newly proposed hybrid algorithm for collision avoidance under different weather conditions. A leading target vehicle brakes abruptly while merging onto a highway, risking collision with the ego vehicle that follows it. A video of this scenario is provided in \cite{youtube_es5}.}
	\vspace{3mm}
	\item {\color{black}\textit{ES6 - High Mobility, Platooning Scenario}: We simulate the newly proposed hybrid algorithm in the dynamic platooning scenario described in \cite[\textit{SM2}]{soner_tvt} and repeat the simulation for different target vehicle speeds (0.25x, 0.5x, 2x and 4x) to quantify the \quotes{mobility-induced error} arising from finite-rate estimation in vehicular VLP performance. }
	\vspace{3mm}
	\item {\color{black}\textit{ES7 - Real Light Patterns and TX-RX Placements}: We repeat the platooning scenario in \textit{ES6} with normal speed, but with different TX light patterns and placements on the target vehicle. Specifically, we evaluate performance under a measured real asymmetric pattern from \cite{memedi_headlightpatterns} as well as Lambertian patterns with different half-power angles (20\textdegree, 35\textdegree and 50\textdegree) to emulate various radiation pattern examples from the literature \cite{becha_ranging, turan_3tx, pw_lambert}. Likewise, we evaluate the system for different TX placements, ranging from on-axis TX-RX placement to a 60cm height offset, similar to the experimental study in \cite{memedi_headlightpatterns}.}
\end{itemize}
\vspace{1mm}

\subsubsection{ES1 - Characterizing Range Measurement Techniques}

\begin{figure}[t]
	\centering
	\includegraphics[width=0.48\textwidth]{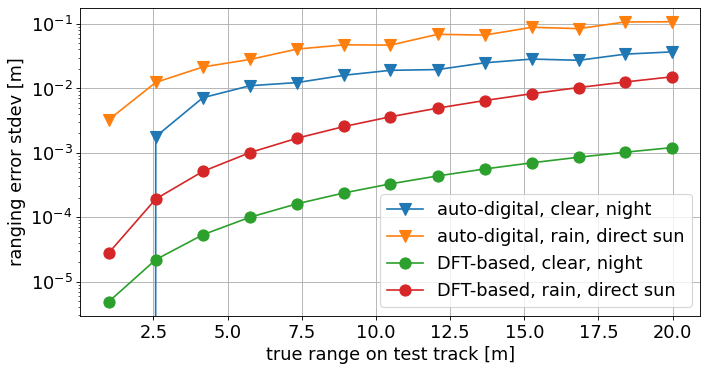}
	\vspace{-2mm}
	\caption{\textit{ES1} - Performance of the two range measurement techniques \cite{becha_ranging, roberts_pdoa} on a 1-20m straight test-track under various weather conditions.}
	\vspace{-2mm}
	\label{ranging}
\end{figure}

Fig. \ref{ranging} shows the sampled standard deviation of the ranging error {\color{black}over 100 iterations} for the two range measurement techniques, i.e., auto-digital \cite{becha_ranging} and DFT-based \cite{roberts_pdoa}, for static target vehicle locations on a 1-20 m straight test track under both favorable and adverse weather conditions. Since the auto-digital technique is susceptible to systematic error due to both quantization and heterodyning \cite{becha_vlr_experimental}, we make two arrangements to observe the effect of channel noise in isolation: 1) The true range values are chosen to be exact multiples of the resolution limit of the auto-digital technique to avoid systematic quantization error (statistical quantization error due to noise will still be present), and 2) each point on the test-track is simulated with a fixed time window that is free of heterodyning error by definition. The auto-digital technique is tuned to $\approx$0.75 cm resolution for 100 Hz measurement rate ($r$=5000, $N$=4, $f_{clock}$=100 MHz with the notation in \cite{becha_ranging}), which results in a $\approx$10 ms simulation time window for each point on the test track. 

Results show that the DFT-based technique is more accurate even in the face of SNR drop throughout the test-track (i.e., with increasing range since TX power is constant) and also with worsening weather conditions. Furthermore, the quantization effect of the auto-digital technique is also seen: At the 1 m true range mark in the clear-night scenario, the output of the auto-digital technique does not change among different samples since none of the noise samples are big enough to shift the output from the same quantization interval, causing standard deviation to be 0 (the 2.5 m mark sample shoots down towards negative infinity, which is virtually where the 1 m sample is on the logarithmic vertical axis). Therefore, in this specific case, the auto-digital technique produces a perfect output, but this only happens because the target vehicle is positioned at an exact multiple of the resolution limit, which is a very scarce occurrence in high-mobility road scenarios. In summary, the relatively more complex DFT-based technique is more accurate than the auto-digital technique against channel noise, making it a better fit for use in vehicular VLP when complexity requirements are not extremely stringent. 

\begin{figure*}[t]
	\centering
	\includegraphics[width=0.95\textwidth]{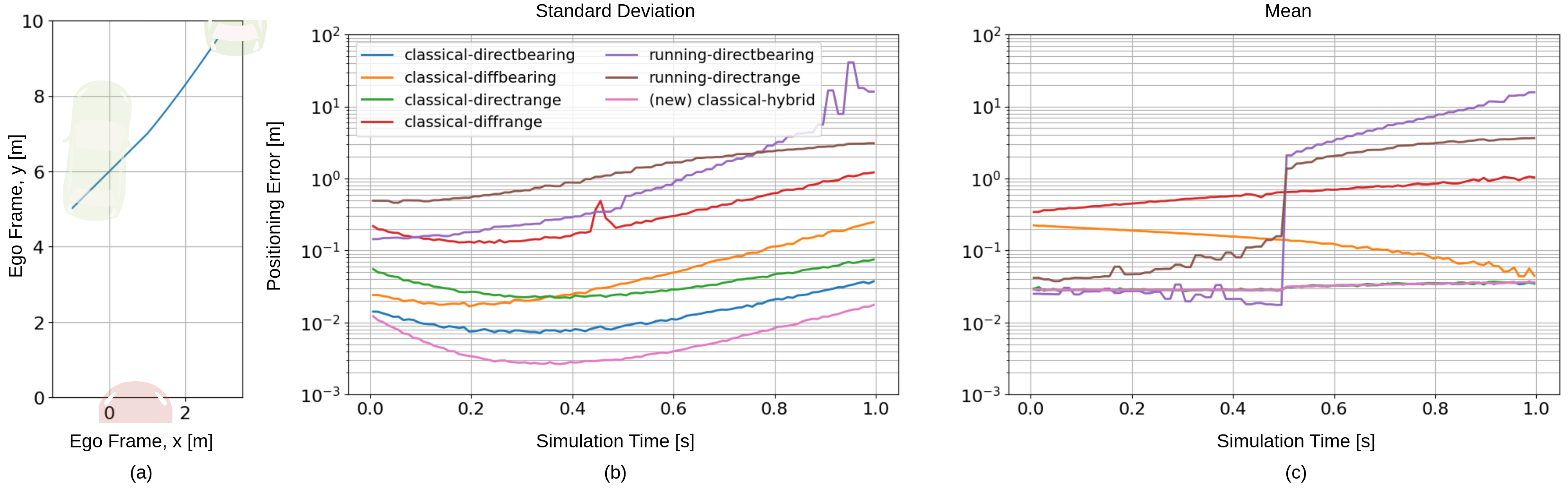}
	\vspace{-1mm}
	\caption{{\color{black}\textit{ES2} - (a) Trajectory of the lane change scenario, (b) standard deviation and (c) mean of positioning error for all six existing vehicular VLP methods listed in Table \ref{algorithms} as well as the novel hybrid method described in Section III-C under day-time, clear weather and indirect sun exposure.}}
	\vspace{0mm}
	\label{benchmark}
\end{figure*}

\vspace{2mm}
{\color{black}\subsubsection{ES2 - Benchmark, All Methods, Lane Change Scenario}

Fig. \ref{benchmark} shows the positioning error for all six existing vehicular VLP methods listed in Table \ref{algorithms} as well as the novel hybrid method described in Section III-C in the lane change scenario described in \cite{techreport_vlp_sonercoleri} under clear day-time indirect sun exposure conditions. Fig. \ref{benchmark}a shows the trajectory, Fig. \ref{benchmark}b shows the sampled standard deviation and Fig. \ref{benchmark}c shows the sampled mean of the positioning error over 1000 iterations of the same simulation run. The results demonstrate that all methods other than those based on classical position fixing and direct parameter measurements suffer significantly in accuracy. The differential measurement based methods have significant bias since they assume target and ego vehicles to be parallel while in reality they aren't, as can be seen in Fig. \ref{benchmark}a and Fig. \ref{benchmark}c. The running fix based methods suffer significantly when the target vehicle starts to change its heading dynamically (i.e., after passing over to the right lane at the 0.5s mark) since they assume relative heading and speed not to change within an estimation interval. The direct bearing/range-based classical position fixing methods as well as the novel hybrid method are the only ones that can sustain cm-level error. Furthermore, the novel hybrid method outperforms the existing state-of-the-art methods using direct bearing/range-based classical position fixing since it has less standard deviation in the positioning error.}

\vspace{2mm}
\subsubsection{{\color{black}ES3 - Validating Theoretical Performance Analyses}}

In this evaluation scenario, the target vehicle is positioned at various locations within a \textless20 m radius of the ego vehicle (i.e., bearing$\neq$0) at longitudinally parallel orientation. At each target vehicle location, the parameter measurements are simulated {\color{black}over 1000 iterations of the noise process}, and the resulting parameter measurement distributions are used for evaluating the CRLB on positioning accuracy for each method. {\color{black}The goal of this simulation is to validate the theoretical analyses in \cite{techreport_vlp_sonercoleri} which claim that all of the classical position fixing algorithms discussed so far converge to the minimum variance unbiased estimator (MVUE) for the parameter measurements they utilize under moderate to high RX SNR conditions.}

\begin{figure}[t]
	\centering
	\includegraphics[width=0.49\textwidth]{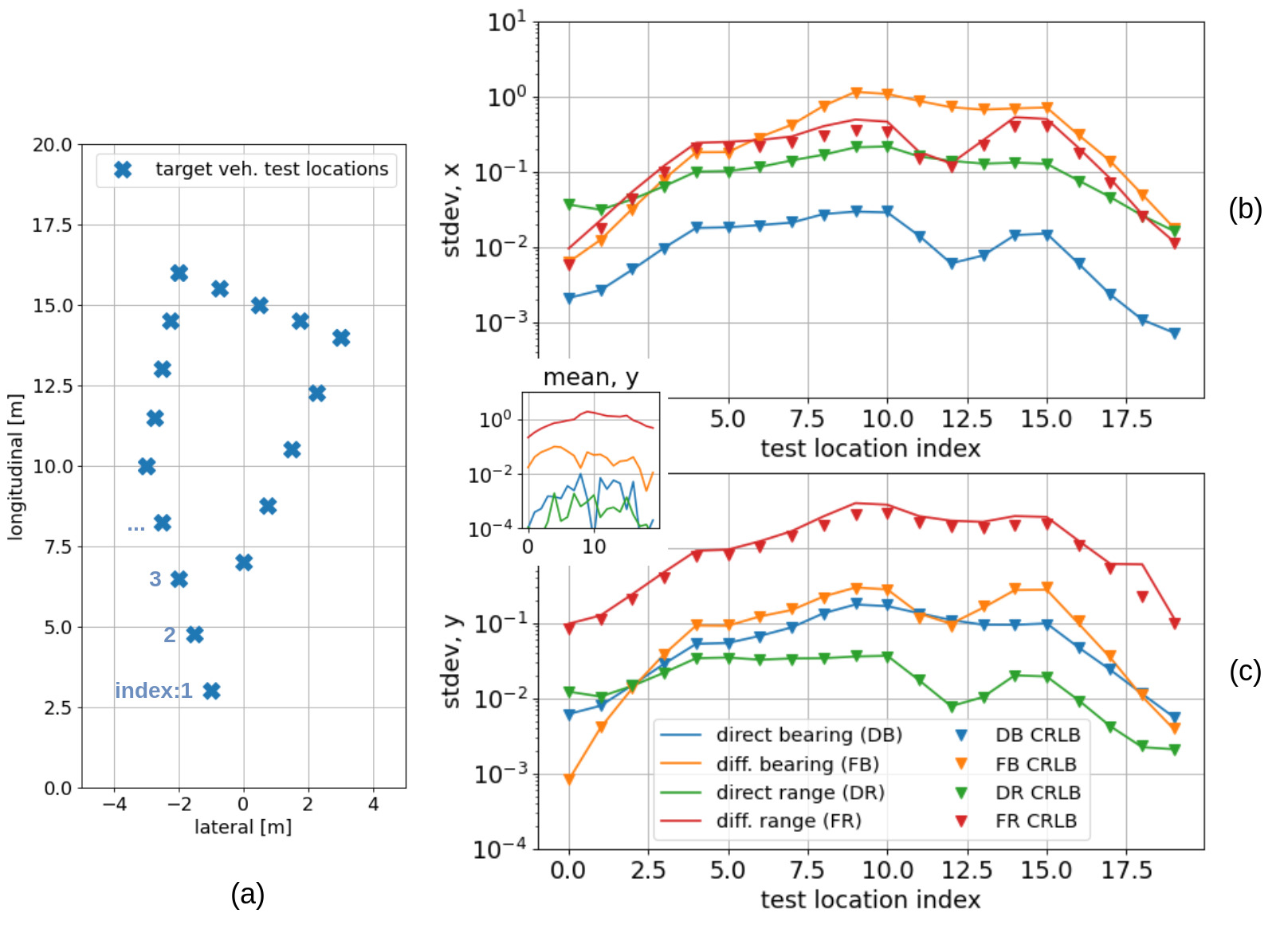}
	\vspace{-1mm}
	\caption{{\color{black}\textit{ES3} - CRLBs and simulated estimation error in the lateral (x) and longitudinal (y) axes for classical fixing methods evaluated for test locations shown in (a), under foggy night time conditions in (b) and (c), respectively. Test location indices start from the bottom-most location shown in (a) and go around the track in clockwise direction.}}
	\vspace{-3mm}
	\label{classical}
\end{figure}

Fig. \ref{classical}a shows the test locations, and Fig. \ref{classical}b and c show the standard deviations as predicted by the CRLB in the lateral ($x$) and longitudinal ($y$) axes, respectively, for {\color{black}foggy weather at night}. The outputs of the algorithms are also sampled to test the MVUE claim in \cite{techreport_vlp_sonercoleri}. The results empirically support this claim since the sampled outputs of the algorithms all approximately match their respective CRLBs. However, {\color{black}as shown in Fig. \ref{classical}c, the extremely low SNR at the farthest points on the track cause deviation from the CRLBs and a significant non-zero bias (shown on the inset) on the differential measurement based methods (as discussed in \cite[Section III-B]{techreport_vlp_sonercoleri}), making them virtually unusable at high noise.} 

In summary, the results show the following: 1) direct measurements provide higher robustness against channel noise, thus, are more suitable than differential measurements for high accuracy in vehicular VLP methods, and {\color{black}2) in support of the results in \textit{ES2}, we observe why the novel hybrid method provides the best performance since} direct range-based methods provide the highest accuracy in the $y$ axis, while direct bearing-based methods do the same in the $x$ axis.

\begin{figure}[t]
	\centering
	\includegraphics[width=0.48\textwidth]{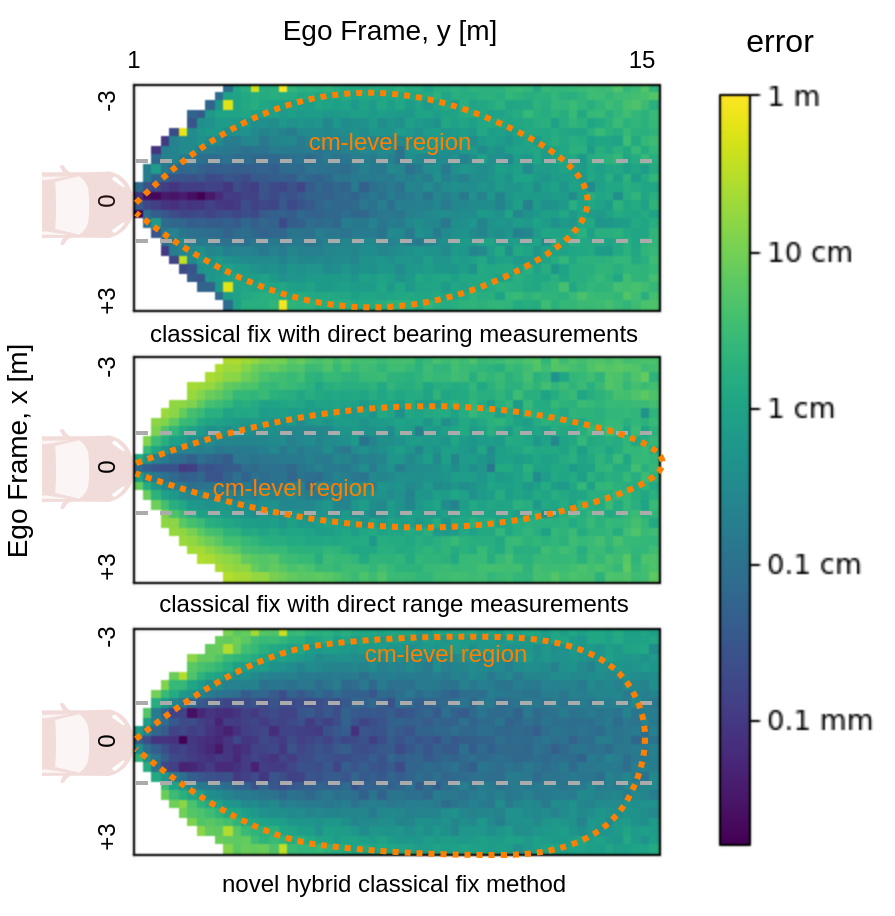}
	\vspace{-1mm}
	\caption{{\color{black}\textit{ES4} - Operational range of the direct measurement based classical fixing methods and the novel hybrid method that combines them, evaluated over a 6 m x 14 m grid covering approximately a 3-lane road. Simulation takes place in daytime, clear weather conditions with indirect sun exposure. White pixels on the grid refer to points where the TX is out of the FoV of the RXs, denoting a loss of estimation.}}
	\vspace{-1mm}
	\label{oprange}
\end{figure}

\vspace{5mm}
{\color{black}
\subsubsection{ES4 - Characterizing Vehicular VLP Operational Range}
Fig. \ref{oprange} lays out a 6 m x 14 m grid over which the sampled standard deviation of positioning error attained by the top three performers from \textit{ES2} are simulated under daytime, clear weather, indirect sun exposure conditions, measured over 50 iterations. From top-to-bottom, the performance of the following methods are shown, respectively: Classical fix with direct bearing measurements \cite{soner_tvt}, classical fix with direct range measurements \cite{becha_positioning}, and the novel hybrid method described in Section III-C which combines the outputs of those two methods. The results verify the original hypothesis of the hybrid method: The bearing-based method performs better on the lateral axis, and the range-based method performs better on the longitudinal axis, as demonstrated by the orange cm-level performance boundaries extending towards the x axis and y axis, respectively. The hybrid method gets the \quotes{best of both worlds} by combining the outputs of the two and covers a much larger region with cm-level accuracy. Furthermore, the results also characterize the cm-level performance range of the novel hybrid method under these conditions to be almost a complete 6 m x 14 m grid, which complies with collision avoidance and platooning requirements.}

\begin{figure*}[t]
	\centering
	\includegraphics[width=0.90\textwidth]{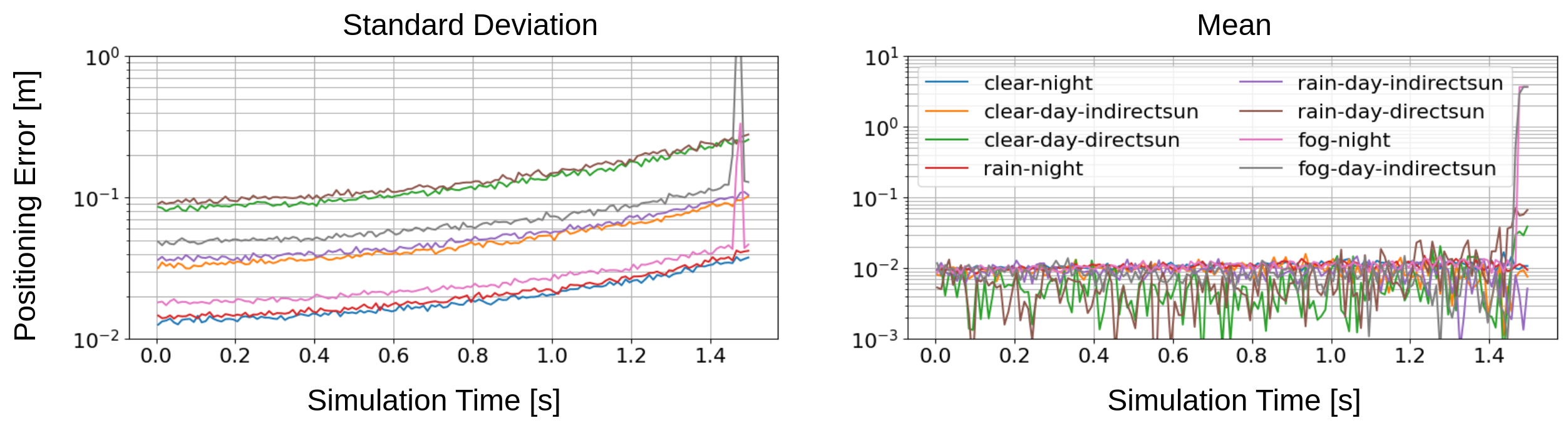}
	\vspace{-1mm}
	\caption{{\color{black}\textit{ES5} - Evaluation of the novel hybrid method in a recorded collision threat scenario from the INTERACTION dataset \cite{interaction_dataset, youtube_es5} under all weather and sunlight-induced shot noise conditions.}}
	\vspace{-1mm}
	\label{colavd}
\end{figure*}

\vspace{2mm}
{\color{black}
\subsubsection{ES5 - Weather Conditions, Collision Avoidance}

Fig. \ref{colavd} shows the results for \textit{ES5}, which simulates the performance of the novel hybrid method in a recorded collision threat scenario extracted from the INTERACTION dataset \cite{interaction_dataset} under all weather conditions considered in this paper, sampling results over 500 iterations of the same trajectory. A video of the scenario showing vehicle trajectories can be found in \cite{youtube_es5}. While the mobility level in this scenario is relatively low (hence, lower mean error), the trajectory here is taken from actual vehicles on a road, which improves the practical relevance of the simulations. Three different weather conditions (clear, rainy and foggy) and three different sunlight conditions (night, daytime indirect sun, and daytime direct sun exposure) are considered to stimulate different shot noise and signal attenuation levels. Attenuation levels for different weather conditions are taken from the experimental results in \cite{fograin} and sunlight-induced shot noise photocurrent level on the RX photodiode for different light exposure conditions are taken from the experimental results in \cite{moreira1997optical}. Foggy weather and direct sunlight are not considered together since that is not a probable natural occurrence. Results show that the increase in shot noise due to sunlight exposure has a larger effect on performance compared to signal attenuation due to weather conditions, and their combined effect can make errors up to eight times higher compared to the best performance under clear, night-time conditions. Nevertheless, cm-level performance is preserved under all adverse conditions except for direct sunlight exposure in this collision avoidance scenario.}

\vspace{2mm}
{\color{black}
\subsubsection{ES6 - High Mobility, Platooning Scenario}
Fig. \ref{platooning} shows the sampled standard deviation and mean of positioning error over 500 iterations for the novel hybrid classical position fixing method during a high-mobility platooning scenario at different target vehicle speed levels under 100 Hz positioning rate like in all other simulation scenarios. The trajectory shown in Fig. \ref{platooning}a is repeated at different speeds by the target vehicle to demonstrate the effect of different levels of mobility on the performance of algorithms. To cover a wide range, we test for the following average relative target vehicle speed levels over the trajectory: 12.31 km/h (4x slow), 24.62 km/h (2x slow), 49.26 km/h (default case), 98.62 km/h (2x fast), 197.63 km/h (4x fast). Although the two fastest cases are extremely improbable since they would require immense torque and handling by the target vehicle (note that the ego vehicle is already moving and these are relative speeds, so the actual speeds need to be even higher), we include them in the simulation to demonstrate the greater than 50 Hz rate positioning rate requirement discussed in Section II.

\begin{figure}[t]
	\centering
	\vspace{-1mm}
	\includegraphics[width=0.41\textwidth]{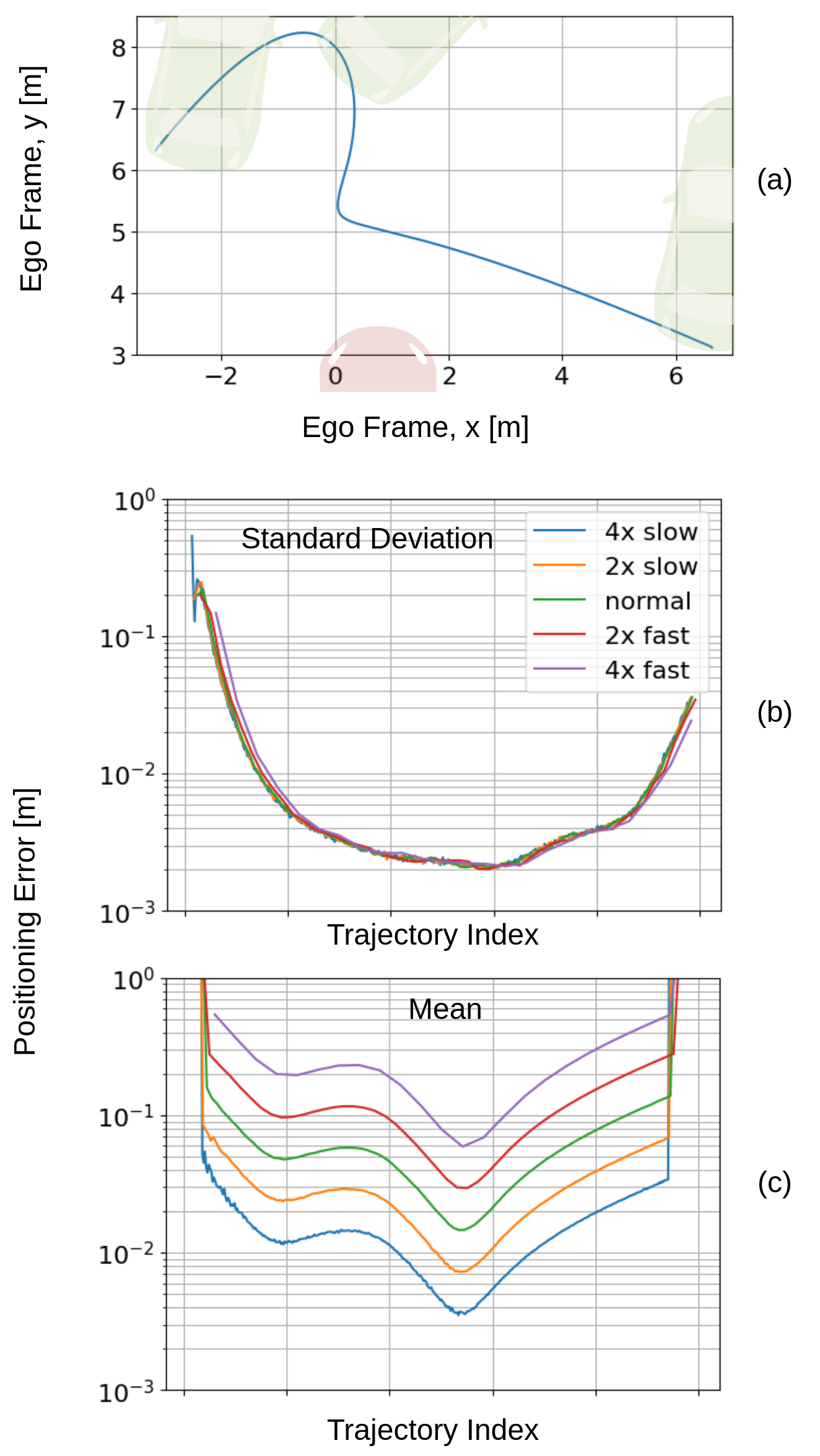}
	\vspace{-1mm}
	\caption{{\color{black}\textit{ES6} - Evaluation of the novel hybrid method in a high-mobility platooning scenario at different target vehicle speeds. Horizontal axis for (b) standard deviation and (c) mean of positioning error shows trajectory location indices rather than time since the same trajectory in (a) is repeated at different speeds, thus, at different time intervals. The non-existent data points at the start and end of the trajectory correspond to sections where the TX is out of the FoV of the RXs, denoting a loss of estimation.}}
	\label{platooning}
	\vspace{-2mm}
\end{figure}

In the mean error plot (Fig. \ref{platooning}c), each point shows the mobility-induced error, which is the maximum error occurring within an estimation interval as described in Section II-B and also in detail in \cite{soner_tvt}. This plot shows the effect of the finite-rate estimations not being able to \quotes{catch up} with the fast-moving target vehicle, which only affects the mean of positioning error since it is deterministic and unknown, i.e., bias. The plot for the standard deviation in Fig. \ref{platooning}b of the error shows the non-deterministic effect of channel noise, and therefore stays relatively similar among different speeds as expected since speed does not affect channel noise directly (note that Doppler effects are found to be insignificant in earlier works \cite{becha_vlcr_experimental}). Results show that except for extremely high mobility cases which are physically improbable (the 2x and 4x fast cases), the novel hybrid method provides cm-level accuracy at the 100 Hz positioning rate simulated in this scenario. This forms the basis of the greater than 50 Hz positioning rate requirement which considers realistic relative vehicle speeds for collision avoidance and platooning applications.}

\vspace{2mm}
{\color{black}
\subsubsection{ES7 - Real Light Patterns and TX-RX Placements}
Fig. \ref{lightandheight} shows the results for the same platooning scenario in \textit{ES6} repeated at normal target vehicle speed, but with different TX light patterns and placements on the target vehicle. Performance is evaluated under four different light patterns and five different TX relative height offsets. For the light patterns, three different Lambertian patterns with half-power angles of 20\textdegree, 35\textdegree ~and 50\textdegree, as well as one real asymmetric pattern experimentally measured in \cite{memedi_headlightpatterns} are tested. The 20\textdegree ~Lambertian is the one used in all of the preceding simulations since it is the closest analytical pattern to taillight regulations \cite{becha_ranging} as well as measured patterns \cite{turan_3tx}. Using the 20\textdegree ~Lambertian, the simulation scenario is repeated at five different TX height offsets to quantify the effect of a headlight height mismatch between the target and ego vehicles. These five different offsets are given by, $z=\{0, 15, 30, 45, 60\}$ cm.

\begin{figure}[t]
	\centering
	\vspace{-1mm}
	\includegraphics[width=0.49\textwidth]{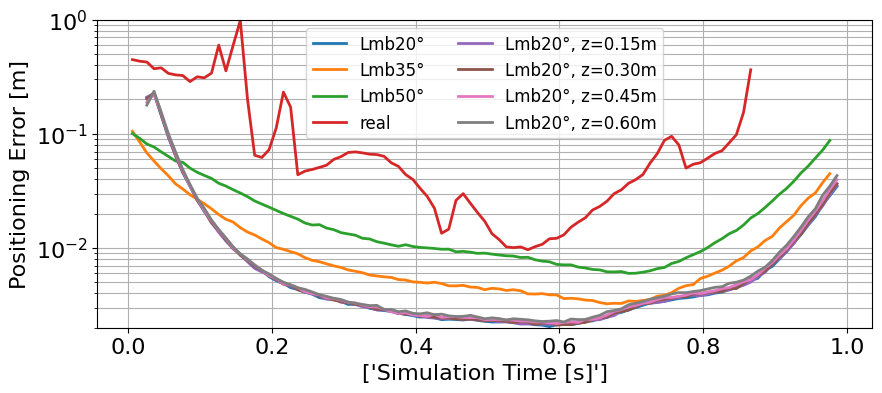}
	\vspace{-1mm}
	\caption{{\color{black}\textit{ES7} - Evaluation of the novel hybrid method by repeating the platooning scenario in \textit{ES6} at normal (1x) speed but with different TX headlight patterns and relative placements (height offsets, or z-levels). Three different Lambertian patterns with half-power angles of 20\textdegree, 35\textdegree ~and 50\textdegree, as well as one real asymmetric pattern experimentally measured in \cite{memedi_headlightpatterns} are tested. Five different height offsets are tested, $z=\{0, 15, 30, 45, 60\} cm$.}}
	\label{lightandheight}
	\vspace{-2mm}
\end{figure}

Results show that while the TX height offset has minimal effect on the performance, the headlight pattern has three significant effects on positioning performance. First, asymmetric and/or non-smooth patterns like the real measured ones from \cite{memedi_headlightpatterns} cause significant variance in positioning performance over a trajectory (shown in red in Fig. \ref{lightandheight}) even if the trajectory itself is smooth as in Fig. \ref{platooning}a. This occurs since the TX beam is asymmetric in both polar axes, causing it to move in and out of the RX many times over the course of the trajectory. Second, the real pattern causes lower overall accuracy compared to the Lambertian patterns. The reason for this is that the real pattern is of a sedan low-beam (LbSedan1, \cite{memedi_headlightpatterns}), which has a downward aim in its main lobe to avoid blinding drivers in oncoming traffic with excessive light, and the target vehicle mostly remains outside this downward-looking main lobe. We chose to use this low-beam as the real pattern to quantify the effects of a severely asymmetric pattern since real taillights heavily resemble Lambertians \cite{turan_3tx} which are already simulated. Third, a wider beam such as the higher half-power angle Lambertians enables the cm-level performance to be preserved over a wider area (i.e., the 0.0 to 0.1s region), but the performance naturally degrades during regions of the trajectory in which the TX beam shines directly on the RX since the power is not as concentrated to the RX as it is in the case with the narrower beam. In summary, the findings match the directional nature of the TX beams well: Wider and smoother beam patterns allow for higher coverage in high-mobility cases, but narrower beams allow for higher accuracy when the TX beam aligns with the RX active area and asymmetric beam patterns cause frequent fluctuations and outage when the beam is not aimed towards the RX.}

\section{Open Questions and Challenges}

In this section, we present remaining important open questions and challenges for the widespread deployment of vehicular VLP under the following categories: 1) further performance improvements and analyses for vehicular VLP, 2) integration of vehicular VLP into the currently evolving \quotes{connected and autonomous vehicle} (CAV) software stack and enabling its widespread deployment in thereof, and 3) co-design of vehicular VLP with sensor systems to enable better localization for upstream applications.

\subsection{VLP Performance Improvements and Further Analyses}

\begin{figure}[t]
	\centering
	\vspace{-1mm}
	\includegraphics[width=0.48\textwidth]{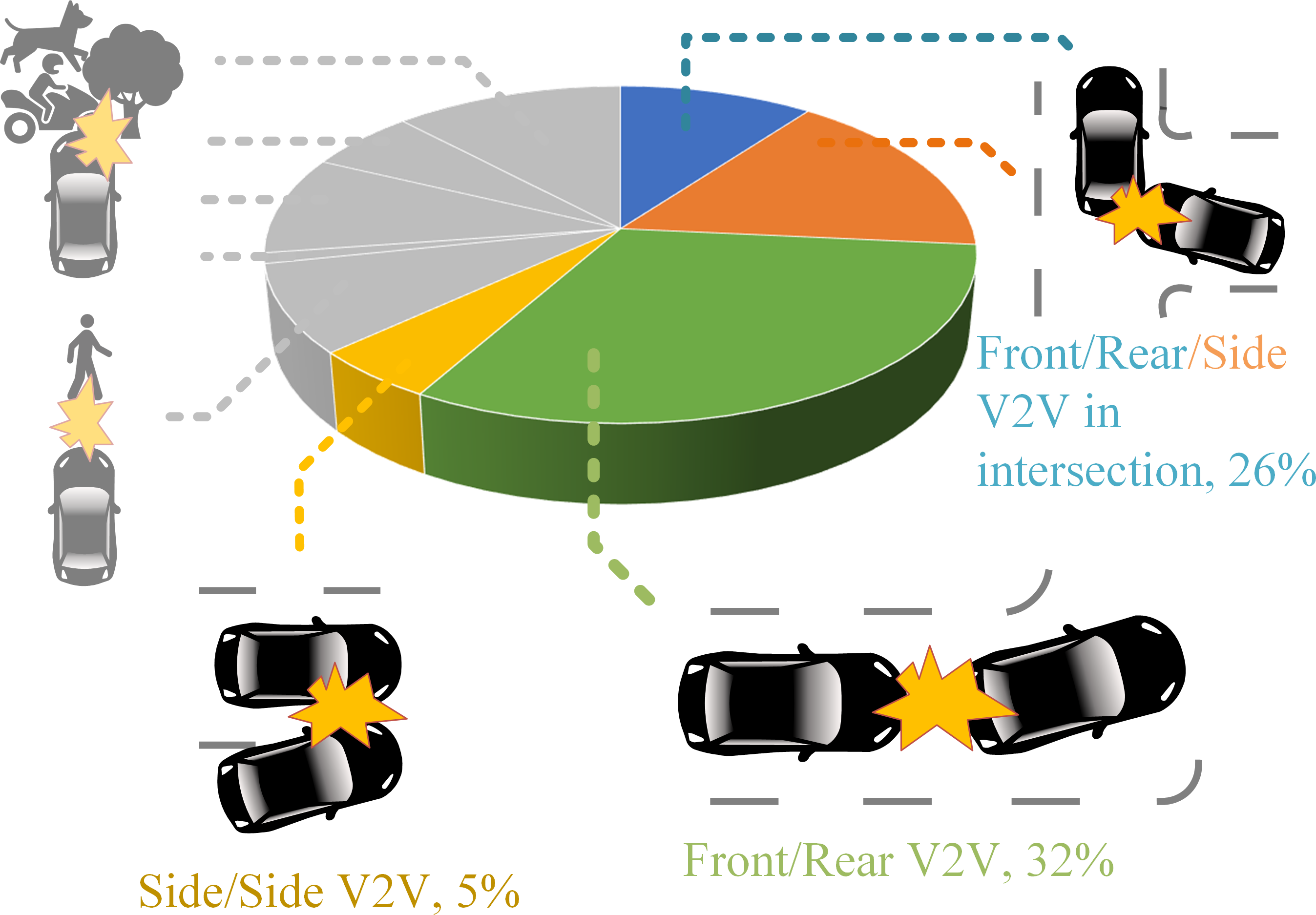}
	\vspace{-1mm}
	\caption{{\color{black} Distribution of traffic accidents based on configuration of victims involved, as measured in the annual \quotes{Verkehrsunfälle} report by Federal
Statistical Office of Germany (DESTATIS) \cite{destatis}}}
	\label{destatis}
	\vspace{1mm}
\end{figure}

One of the main shortcomings of current vehicular VLP methods is insufficient FoV: The current setups that consider only front/rear-facing VLC units {\color{black}can only cover half of all vehicle-to-vehicle collisions as shown in Fig. \ref{destatis} since they} cannot detect pure side collision threats \cite{kaempchen, destatis}. Although existing methods can be adapted for detecting such collisions via additional RX units on the sides \cite{uysal_rx_placement}, the effectiveness and regulation compliance of such configurations are yet to be investigated. Moreover, current classical position fixing methods have dead zones within their FoV limiting their coverage: target TXs need to be inside the FoVs of both RXs at the same time for a fix. Therefore, developing better running position fixing methods or hybrid bearing/range classical methods that use a single RX (similar to radial and distance measurement equipment (DMEs) in aviation \cite{dme_vor}) are important directions for improving vehicular VLP FoV. 

Another important line of research is further improving the accuracy and rate of vehicular VLP methods: Using an extended set of measurements by adding more RX units or taking more samples over time and solving over-determined systems of equations (i.e., typically via least-squares or machine learning) is a promising research direction, and has recently seen some elementary progress \cite{xu_matchedfilt_lm}. Other promising research directions include new RX optical and electrical configurations \cite{cailean_noiseadaptive} and parameter measurement techniques (e.g., via positioning-motivated visible light waveform design \cite{vlp_optimalpulse}) with higher robustness against noise, and further \quotes{environment-aware} methods that sense and adapt to channel conditions for higher accuracy \cite{cailean_envradaptive}. The theoretical performance of such new techniques can be investigated via the generalized bounds derived in \cite{mfkeskin_bounds}, and the positioning performance they enable can be evaluated via the CRLBs derived in \cite{techreport_vlp_sonercoleri}. 

Lastly, we note that experimental studies in this area are currently very limited. {\color{black}Notable experimental studies from the vehicular VLP field are \cite{becha_vlcr_experimental, becha_vlr_experimental} for the auto-digital ranging method described in Section II-A, but these are laboratory experiments in controlled environments rather than field experiments, and other methods in the field (e.g., bearing-based) still lack thorough experimental validation. Field experiments have been more popular for vehicular VLC since precision and repeatability in vehicle placement and movement is not as critical as in VLP in those studies \cite{vlc_channel_wnl, vlc_channel_wnl_ml, aly2020experimental, eso2019experimental, aly2021experimental}. These studies have helped in characterizing the vehicular VLC channel, which indirectly also contributed to vehicular VLP since the two share the same physical medium.} However, although theoretical and simulated analyses done so far have provided a good basis for understanding vehicular VLP performance in realistic scenarios {\color{black}by enabling investigation into field configurations that would have been hard to investigate experimentally,} comprehensive experimental studies on the road are still necessary for fully characterizing vehicular VLP performance with actual regulation-compliant headlights \cite{memedi_headlightpatterns, memedi_headlightpatterns2}.

\subsection{Integration and Widespread Deployment of Vehicular VLP}

The integration of communication-based positioning methods into the vehicular domain is still an open field. Unlike RF \cite{NAlam_cooperative}, VLP can provide the required accuracy for collision avoidance and platooning applications, but on-vehicle VLC TX and RX units are not ubiquitous yet. {\color{black}Even with ubiquitous deployment, analyses need to be conducted as to the availability of the applied methods, especially regarding daytime operation since vehicle lights are not promised to stay on all the time during the day. Such analyses need to deduce service schedules for vehicular VLP methods based on how long the vehicle lights stay on, and accordingly measure the overall availability of the vehicular VLP service as a suitable complementary method for other on-board sensor-based localization technologies}. Moreover, the position of VLC technologies within the automotive software stack is not clear yet since the stack itself is currently evolving to accommodate \quotes{V2X} connectivity and autonomy \cite{mckinsey_automotivesw}. VLP could either be a standalone localization service or it could exist as an extension of the VLC service without compromising communication functionality \cite{revistedi_vlcp} within these lines. Further research on this aspect is necessary since this choice has implications for the newly developed safety-relevant automotive software standards (e.g., ISO/PAS 21448, SOTIF \cite{sotif}) as well as for future-proofing \cite{rfconvergence}.

One other related challenge, which might be regarded as the holy grail in this area, is attaining modulation-free VLP. Modulation-free VLP considers the detection of random LED intensity fluctuations, which occur due to various random noise factors, such as power supply switching noise. Once the LED is detected, range and/or angle measurements can be done, and the vehicular VLP methods discussed in this paper can be applied for positioning. Since this would enable the localization of \quotes{legacy} vehicles that do not carry dedicated VLC TX modulators, it would vastly improve the adoption speed of vehicular VLP in the automotive sector. While there have been indoor VLP works towards this direction primarily by utilizing so-called \quotes{signals of opportunity} \cite{unmodulated}, this highly valuable technology is still an open question for vehicular VLP.

\subsection{VLP-Sensor Co-design and Upstream Applications}

Vehicular VLP and sensor-based systems can also be jointly designed to exploit localization performance gains from both sides since they have \quotes{orthogonal} performance attributes. For example, VLP position estimations can be used as a new grounding modality for improving the performance of vision-based target trackers \cite{guney_cvav} and trajectory prediction models \cite{ngsim_nn}. Since current examples of such systems typically do grounding by bootstrapping on the same modality (e.g., they utilize \textit{a priori} known features from the same source image for grounding \cite{visualgrounding}) the complexity increases and the rate decreases. These systems could significantly benefit from a new modality like VLP that is low complexity, high rate and high accuracy. 

Similarly, VLP-based collision avoidance and platooning systems could also benefit from sensor systems: The results in this paper have shown that one of the main sources of error in vehicular VLP is due to the estimations not being able to catch up with the high-mobility target vehicle. Since this component changes relatively slowly compared to sensor rates as explained in \cite{soner_tvt}, inertial and visual sensor-based probabilistic trajectory prediction methods (e.g., from SLAM literature \cite{bertha_drive, scanslam, slam_pt2}) can be used for mitigating this mobility-induced error in VLP. The investigation of the collaboration between such sensor-based methods and vehicular VLP methods for improving localization performance in autonomous driving, is a highly promising future research direction.

\section{Conclusion}
This survey paper reviews the state-of-the-art in vehicular VLP, {\color{black}proposes a new method that advances the state-of-the-art}, analyzes the performances of all methods against challenging and realistic road scenarios {\color{black}as well as practical non-idealities}, and discusses future challenges faced for the widespread deployment of vehicular VLP. {\color{black}The proposed novel hybrid classical position fixing method outperforms all existing methods in all considered conditions by combining direct range measurements for longitudinal and direct bearing measurements for lateral coordinate estimations with a receiver that can do both simultaneously. Simulation results show that this method can satisfy the accuracy and rate requirements of collision avoidance and platooning applications under nearly all weather, noise, mobility and light pattern and placement conditions, proving its eligibility.}

Open challenges for vehicular VLP research includes environment and noise-adaptive operation and further analyses of performance under different use cases, including, {\color{black}first and foremost}, comprehensive experimental studies. Furthermore, integration of vehicular VLP as a relative vehicle localization technology into the modern CAV software stack is an important topic that requires attention from both the academia and the industry due to its implications for standardization of safe autonomous driving. Similarly, practical challenges such as increasing the FoV of vehicular VLP methods to enable side collision detection, and finding new signal-of-opportunity based target detection methods that do not require explicit LED modulation to enable positioning of \quotes{legacy} vehicles without VLC hardware, require further development efforts. 

Solutions to these problems and challenges would free vehicular VLP of its technical restrictions and induct it into the vehicular localization framework alongside sensor-based methods. This would pave the way for the ultimate goal in vehicular VLP, which is complementing and improving sensor-based vehicle localization systems for fully autonomous collision avoidance and platooning applications.



\bibliographystyle{IEEEtran}
\bibliography{vlp_survey}

\end{document}